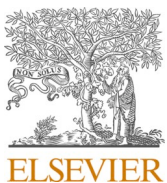
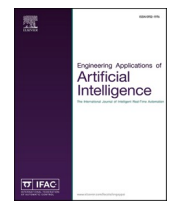

# Comparison of edge computing methods in Internet of Things architectures for efficient estimation of indoor environmental parameters with Machine Learning

Jose-Carlos Gamazo-Real [a],[*], Raúl Torres Fernández [b], Adrián Murillo Armas [b]

[a] *Universidad Politécnica de Madrid. Information Systems Department, Computer Architecture and Technology Area, ETSI Sistemas Informáticos, 28031, Madrid, Spain*
[b] *Universidad Politécnica de Madrid, 28040, Madrid, Spain*

A B S T R A C T

The large increase in the number of Internet of Things (IoT) devices have revolutionised the way data is processed, which added to the current trend from cloud to edge computing has resulted in the need for efficient and reliable data processing near the data sources using energy-efficient devices. Two methods based on low-cost edge-IoT architectures are proposed to implement lightweight Machine Learning (ML) models that estimate indoor environmental quality (IEQ) parameters, such as Artificial Neural Networks of Multilayer Perceptron type. Their implementation is based on centralised and distributed parallel IoT architectures, connected via wireless, which share commercial off-the-self modules for data acquisition and sensing, such as sensors for temperature, humidity, illuminance, $CO_2$, and other gases. The centralised method uses a Graphics Processing Unit and the Message Queuing Telemetry Transport protocol, but the distributed method utilises low-performance ARM-based devices and the Message Passing Interface protocol. Although multiple IEQ parameters are measured, the training and testing of ML models is accomplished with experiments focused on small temperature and illuminance datasets to reduce data processing load, obtained from sudden spikes, square profiles and sawteeth test cases. The results show a high estimation performance with F-score and Accuracy values close to 0.95, and an almost theorical Speedup with a reduction in power consumption close to 37% in the distributed parallel approach. In addition, similar or slightly better performance is achieved compared to equivalent IoT architectures from related research, but error reduction of 35–76% is accomplished with an adequate balance between performance and energy efficiency.

## 1. Introduction

For over the last few years, the technologies and devices related to Internet of Things (IoT) have quickly gained popularity due to their sensor-fusion property for aggregating great volumes of data from multiple data sources. In 2017, total spending on these technologies reached nearly $2 trillion and 20 billion of connected devices were forecasted by 2020 (Gartner Inc., 2017) and, in 2023, the global edge computing market is expected to reach $1.12 trillion (Walker et al., 2022). In addition, it is estimated that close to 50 million devices will be connected by 2030, forcing IoT architectures to be more scalable, efficient and autonomous (Gloria and Sebastiao, 2021). The IoT paradigm shift from cloud to edge has contributed to this evolution, placing increasing importance on computing close to physical data sources using resource-constrained devices (Samie et al., 2019), referred to as edge computing. However, the growing number of applications that require forecasts in a short period of time encounter barriers with the reduced performance of IoT devices with respect to power consumption, processing, and connectivity (Shafique et al., 2018). In addition, the problem of limited resources is accentuated with the incorporation of estimation methods based on Artificial Intelligence (AI) directly embedded in IoT edge devices, such as Artificial Neural Networks (ANN) (Alam et al., 2017).

The edge computing introduces a new computation layer physically closer to the end-users that tries to overcome the problems of cloud computing, such as communication latencies, network congestion, and security issues (Harish et al., 2020) in applications of computer vision, speech recognition, natural language processing or weather forecast. In such intelligent applications, the communication and processing are the

**Abbreviations**

| | |
|---|---|
| AI | Artificial Intelligence |
| ADAGrad | Adaptive Gradient Algorithm |
| ADC | Analog-to-Digital Converter |
| AMQP | Advanced Message Queuing Protocol |
| ANN | Artificial Neural Network |
| ARM | Advanced RISC Machine |
| BLE | Bluetooth Low Energy |
| CoAP | Constrained Application Protocol |
| COTS | Commercial Off-The-Self |
| CPS | Cyber-Physical System |
| DT | Decision Tree |
| ELM | Extreme Learning Machine |
| XGB DT | Extreme Gradient Boost DT |
| GPU | Graphics Processing Unit |
| GPUEX | GPU External |
| HPM | High-performance Processing Module |
| HVAC | Heating, Ventilation, and Air Conditioning |
| I$^2$C | Inter-Integrated Circuit |
| IAQ | Indoor Air Quality |
| IEQ | Indoor Environmental Quality |
| IoT | Internet of Things |
| kNN | k-Nearest Neighbours |
| LDR | Light Dependent Resistor |
| LED | Light Emitting Diode |
| LPM | Low-performance Processing Module |
| LTSM | Long Short-term Memory |
| MAE | Mean Absolute Error |
| MAPE | Mean Absolute Percentage Error |
| ML | Machine Learning |
| MLP | Multilayer Perceptron |
| MOX | Metal Oxide |
| MPI | Message Passing Interface |
| MQTT | Message Queuing Telemetry Transport |
| RASPEX | Raspberry External |
| RF | Random Forest |
| RMSE | Root Mean Square Error |
| SBC | Single Board Computer |
| SDG | Sustainable Development Goal |
| SENSAIR | Sensing Air |
| SENSCUR | Sensing Current |
| SENSLUX | Sensing Lux/Illuminance |
| RH | Relative Humidity |
| RMSProp | Root Mean Square Propagation |
| SoC | System-on-a-chip |
| SVM | Support Vector Machine |
| TA | Training Accuracy |
| TF | TensorFlow |
| VOC | Volatile Organic Compound |
| WLAN | Wireless Local Area Network |
| WHO | World Health Organization |
| WSN | Wireless Sensor Network |

major components, but the power consumption is the most limiting factor in edge implementations that support AI processing techniques (Ai et al., 2018; Gloria and Sebastiao, 2021). These solutions are traditionally implemented using a centralised approach, in which IoT sensing nodes collect raw data to be streamed to a concentrator that acts as a gateway, and a remote cloud or another edge device with high processing capacity performs compute-intensive tasks, such as AI model training and inference (Ramos et al., 2019). Although centralisation has the advantage of storing the results of the inference stage and can be later requested by other applications without re-running the computation (Campolo et al., 2021), the vast use of certain central edge nodes with parallelism degree and multiple convolution cores, such as Graphics Processing Units (GPU), are not appropriate for resource-constrained IoT architectures (Shafique et al., 2018). Given this situation, other alternatives are being evaluated with the direct intervention of local nodes in data processing, such as concurrent computing (Shi et al., 2016) and fog computing, acting as a medium layer with fog devices of limited storage and computing capabilities, which collect the data offloaded by IoT sensing devices and decide whether to process it locally or offload to the cloud layer via a fog gateway (Ogundoyin and Kamil, 2023).

Within the AI techniques, a large number of topologies are studied to estimate parameters in cyber-physical systems (CPS) in which edge-IoT architectures could be applied to estimate system parameters, but those based on statistics, regression, classification, clustering, and meta-heuristic optimization techniques stand out above the rest to reduce the negative effect of the parametric uncertainties (Tien et al., 2022; Wei et al., 2019). Examples of CPS using AI techniques for parameter estimation that could be implemented in edge-IoT architectures are electric machines, such as ANNs trained with rotor stages in brushless DC motors (Gamazo-Real et al., 2022), Black Widow Optimization (BWO) in membrane fuel cells to meet energy needs with reliability (Singla et al., 2021), and an Equilibrium Optimizer (EO) enhanced with an ANN in photovoltaic cells to obtain maximum output energy harvest (Wang et al., 2021). As these different examples illustrate, there is a concern for energy balance in a wide variety of smart CPS (Muhuri et al., 2019), which together with the growing interest in applications related to the monitoring of air quality for enhanced living environments (Marques et al., 2020), both indoors and outdoors, has motivated the application of AI techniques to achieve responsible sustainable environments with respect to the energy consumption (Alam et al., 2017). This aspect currently has a great relevance according to United Nations Sustainable Development Goals (SDG), such as the SDG 11 on sustainable cities and communities, SDG 3 on good health and well-being, and SDG 3 on affordable and clean energy (United Nations, 2023).

Within CPS related to sustainable environments, the applications associated to indoor environment quality (IEQ) have a great relevance since they directly impact the comfort levels of a building, considering that buildings contribute almost 40% of total energy consumption in developed countries (Pérez-Lombard et al., 2008), and the indoor air pollution is the leading cause of millions of serious premature sicknesses (Elnaklah et al., 2021). Generally, the IEQ evaluation does not assess the overall indoor environment, but rather assess the aspects of indoor environment separately, such as thermal, acoustic, visual and indoor air quality (IAQ) (Wei et al., 2020) as the principal factors by the ASHRAE standard (ASHRAE, 2023). The World Health Organization (WHO) has made frequent efforts to improve and refine air quality standards from the definition of air quality guidelines on pollutants in 2005 (WHO, 2006). Lack of model comparison is a major obstacle to assessing the accuracy and robustness of existing models (Tang et al., 2022), as some approaches that are widely used in predicting IEQ and building energy assumption, such as Machine Learning (ML) due to the ability to identify and learn underlying patterns in massive data with almost no human intervention (Tian et al., 2021). However, they could have validation gaps on separate data not used in the model development to ensure that the model does not overfit and then fail to fit the new data (Wei et al., 2019). At present, the use of ML has spread in the field of energy efficiency and indoor environment analysis in buildings (Copiaco et al., 2023), and although there are numerous algorithms to carry out the learning process, the most widely used are the ANNs of type Multilayer

Perceptron (MLP) for their ability to model nonlinear functions between inputs and outputs with accuracy, fault tolerance, and flexibility, due to its extensive interconnectivity (Martínez-Comesaña et al., 2021). Other ML algorithms are also used in the context of indoors, such as Random Forest (RF) that is based on individual regression trees not pruned (Tang et al., 2022; Yu et al., 2021), Support Vector Machine (SVM) which attempts to minimise an upper limit on the generalisation error rather than minimising the prediction error (Leong et al., 2020; Zhong et al., 2019), and decision trees based on Gradient Boost (GBDT) (Almalawi et al., 2022; Moursi et al., 2021).

A relevant point to consider in the implementation of ML estimation models in IoT is the architecture structure with hardware devices, networking standards and communication protocols. To develop a suitable IoT implementation, it is necessary to examine which factors affect the problem in more detail and how stable and economical the models are in real life (Gangwar et al., 2023). This perspective has a special emphasis on the cost of IoT solutions that are energy efficient and scalable to enable rapid changes to the sensing network without additional infrastructure requirements (Scislo and Szczepanik-Scislo, 2021), with the additional great challenge that is the use of low-cost and commercial off-the-self (COTS) solutions to implement the measurement setup. An IoT sensing architecture is usually based on a device that collects data from sensors via wired or wireless network (WSN) and, depending on the structure of the IoT architecture, send it to processors that implement data processing at the edge or forwards it, acting as a gateway, to cloud or fog devices (Ahmed et al., 2020; Mumtaz et al., 2021). Regarding the processors used in edge-IoT nodes for air quality, the concept of single-board computer (SBC) has been widely spread, and they are usually implemented with Raspberry devices (Pi 3 and Pi 4 models, typically) based on ARM processors (Kiruthika and Umamakeswari, 2018; Kumar and Jasuja, 2017; Zhang et al., 2021), Arduino devices (Uno model, typically) based on Atmel microcontrollers (Abraham and Li, 2014; Firdhous et al., 2017; Karami et al., 2018), system-on-a-chip (SoC) devices such as ESP32 (Nasution et al., 2020; Taştan and Gökozan, 2019) and ESP8266/NodeMCU (Siva Nagendra Reddy et al., 2018), and integrated wireless sensor modules (IWSM) that embed in the same board a low-cost microcontroller, a RF circuit and some sensors (Kim et al., 2017). In other IoT implementations, more powerful processing devices are added to the edge to reduce computing times in the ML model implementation, such as GPUs, in which the Nvidia manufacturer has a clear dominance with models such as Jetson Nano (Shah et al., 2020) and GeForce RTX (Cosoli et al., 2022).

The other main components of an IoT architecture are the networking standards and protocols implemented on the processor board, and it can vary depending on the needs of collecting data from sensors or sending it to cloud services. Typically, the network side is based on IEEE 802.11 (Wi-Fi) to use pre-existing infrastructure (Troncoso-Pastoriza et al., 2022), ZigBee using XBee modules (Tsang et al., 2016; Wu et al., 2017), Bluetooth and the version of low power consumption (BLE) (Firdhous et al., 2017), and Wi-SUN (IEEE 802.15.4) to achieve a more stable RF link than the 2.4 GHz-ISM band in WSN deployments (Kim et al., 2017). Over these networking standards, the "thing-driven" communication protocols play an important role in interconnecting sensors to edge processing devices and gateways to cloud services, such as the publish/subscribe protocol Message Queuing Telemetry Transport (MQTT) (Kumar and Jasuja, 2017), the Constrained Application Protocol (CoAP) as a lightweight RESTful protocol based on UDP that inherits the same client/server paradigm adopted in HTTP (Tanganelli et al., 2015), lighter versions of CoAP such as the Lightweight Messaging Protocol (LiMP) (Agyemang et al., 2022), and other approaches such as WebSockets, REST API over HTTP/HTTPS and Extensible Messaging and Presence Protocol (XMPP) (Fioccola et al., 2016). These standards and protocols, together with the ML models, influence the energy efficiency of the architecture, which can be treated as a problem of network power consumption (Alam et al., 2017). Some AI approaches can assist edge-IoT computing to reduce network latencies or congestion (Veeramanikandan et al., 2020) and therefore improve energy efficiency, where the software programming platform can also play a very relevant role. Although there is a great variety in programming languages and software toolkits, it is worth mentioning some of the most used to implement ML models on processing devices, such as Python and R languages with many open-source libraries (Kadiyala and Kumar, 2017; Tang et al., 2022), toolkits such as Mathworks Statistics and Machine Learning toolbox and Deep Learning toolbox for Matlab (Cho and Moon, 2022), and software environments such as Mathworks ThingSpeak (Scislo and Szczepanik-Scislo, 2021).

With these approaches, this article presents the performance assessment of two estimations methods based on different edge-IoT architectures, and a series of experiments on the estimation of environmental parameters are carried out, also considering the power consumption to determine their validity. A distributed parallel architecture is compared with a centralised one, both based on COTS, low-cost circuitry and energy-constrained devices connected via IEEE 802.11 or WLAN (Wireless Local Area Network), implemented with C and Python languages. To compare both proposals, several MLP-ANN topologies (Samie et al., 2019) are implemented with the aim of being simple and lightweight ML models, also applying some general optimization strategies based on TinyML (Raha et al., 2021; Sanchez-Iborra and Skarmeta, 2020). The method based on the centralised architecture is implemented with a Nvidia Jetson GPU, as a powerful central node, to learn the ML models and collect sensor data from devices with fewer processing and memory resources based on ARM processors, such as those embedded in Raspberry Pi boards, that communicate over the network using the MQTT publish/subscribe protocol. Faced with this method, an alternative based on a distributed architecture is used, without a central node, in which the learning of ML models is developed using parallel techniques with ARM-based devices of limited resources that communicate through the Message Passing Interface (MPI) protocol. The experimental phase is carried out in an indoor environment, acquiring and processing temperature, relative humidity and multi-gas parameters, but focused on temperature estimation as a specific test case. In the experiments, the comparison of both methods not only uses classification metrics such as F-score, Accuracy, and Training Accuracy (TA) based on small datasets, but other more IEQ-related metrics, such as the Root Mean Square Error (RMSE) and Mean Absolute Error (MAE), have also been used to compare them with related research. As an advance, it is expected that the method based on a distributed parallel architecture with more limited resources may have an advantage in estimating environmental parameters with sufficient accuracy and lower power consumption.

The main contributions of this article can be summarised as follows.

- Implementation of ML models from a lightweight perspective of resource usage, such as MLP-ANN, using low-cost sensors and COTS processing devices.
- Deployment of edge-IoT architectures from centralised and distributed approaches using devices with different processing capabilities, such as GPUs and ARM-based processors, and various communication protocols, such as MQTT and MPI.
- Estimation performance and energy efficiency of the proposed methods with experimental tests using classification and parallel computing metrics, taking temperature estimation in a real indoor environment as a specific case study.
- Comparative analysis with related research works regarding estimation performance, considering mainly similar edge-IoT architectures for estimating air quality parameters in indoor environments, but also other IoT architectures, such as cloud-IoT and fog-IoT, or single-computer simulations to get a broader perspective.

The remainder of this article is structured as follows. In Section 2, the setup of the experimental prototype to conduct the experiments is explained, followed by the description of the sensor data acquisition with its electronic modules and the format of the data frames to encapsulate this sensor data. This section also considers the software components used in the design, such as programming languages and libraries, and the database model used for local and global data storage. Then, in Section 3, the general estimation methodology is described and, more specifically, the two experimental methods developed for the estimation of environmental parameters together with the evaluation guidelines and the training and testing processes of the ML models based on MLP-ANN. In Section 4, the performance of the methods is validated by actual experiments focusing on temperature and illuminance parameters related to power consumption data, and the results are evaluated and discussed with respect to related research. Finally, conclusions are drawn in Section 5, and Appendices A and B included at the end of the article provide additional details on the experimental setup design and estimation methods and results.

## 2. Experimental setup and data acquisition

In this section, an overview of the setup used in the experimental phase is provided. The implementation of the IoT architectures and sensor data acquisition are analysed in detail as the main components to collect data for training and testing of the proposed methods.

### 2.1. Overview of the prototype setup

The assessment of the methodology for parameter estimation was carried out by conducting several experiments with a specific setup that enabled the implementation of both architectures and collected data. The prototype setup consisted of the processing hardware and learning software to support the test conditions of real environmental data using sensor fusion. Fig. 1 depicts the schematic layouts of the proposed IoT architectures. Fig. 1-a shows the centralised architecture implemented with a high-performance processing module (HPM) based on a Nvidia Jetson Xavier NX GPU board in which the ML algorithms were implemented. The Amper ASL-26555 WLAN router 802.11 b/g/n was used to connect multiple low-performance processing modules (LPM) based on Raspberry Pi 3 B boards through the lightweight publish/subscribe MQTT protocol. LPMs collected environmental data from sensor data acquisition modules, called Raspberry External (RASPEX), which was encapsulated in frames for transmission and storage in the HPM using the GPL-licensed MySQL-derived database MariaDB. The distributed parallel architecture of Fig. 1-b does not include a central node, so the processing was performed in parallel within LPMs using local MariaDB databases and data was exchanged through the MPI protocol. For reference only, an IoT edge node was considered the combination of an LPM and a RASPEX module, which included light sensors (SENSLUX) based on a light dependent resistor (LDR) with sensitivity in the visible light region, air quality sensors (SENSAIR) to measure temperature, relative humidity (RH), $CO_2$ and volatile organic compounds (VOC), and current sensors (SENSCUR) based on a shunt resistance and a differential amplifier to obtain the power consumption of the processing modules. The actual illustration of the experimental setup is shown in Fig. 2.

The main characteristics of LPM and HPM devices that are considered for their selection are listed in Table A.1 of Appendix A, such as the type of processor and processing cores, RAM and cache memories, and the standard of the WLAN interface.

### 2.2. Sensor data acquisition

The acquisition stage played a prominent role by providing sensor fusion data with the appropriate periodicity and ranges in order to be stored in the databases. This operation was implemented using the Raspberry Pi 3 B and RASPEX boards, which were specifically designed for the setup and were based on the MCP3008 10-bit analog-to-digital converter (ADC). As shown in Fig. 3, RASPEX boards included ports for connecting external sensors, such as SENSAIR and SENSCUR, and also a SENSLUX sensor and other ancillary electronics, such as a push button and light-emitting diodes (LED) to control and monitor acquisition status. In addition, due to the need of monitoring the power consumption of the GPU, an external board called GPU External (GPUEX) was also specifically designed for the setup. It was inserted between the DC power supply and the GPU board to measure the current consumption of the GPU by means of a SENSCUR module. The main characteristics of the sensor modules are shown in Table A.2 of Appendix A.

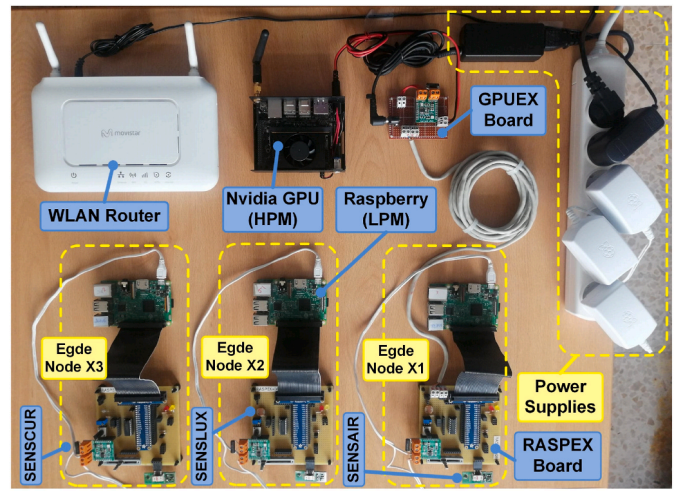

**Fig. 2.** Setup of the complete prototype setup used in the experimental phase.

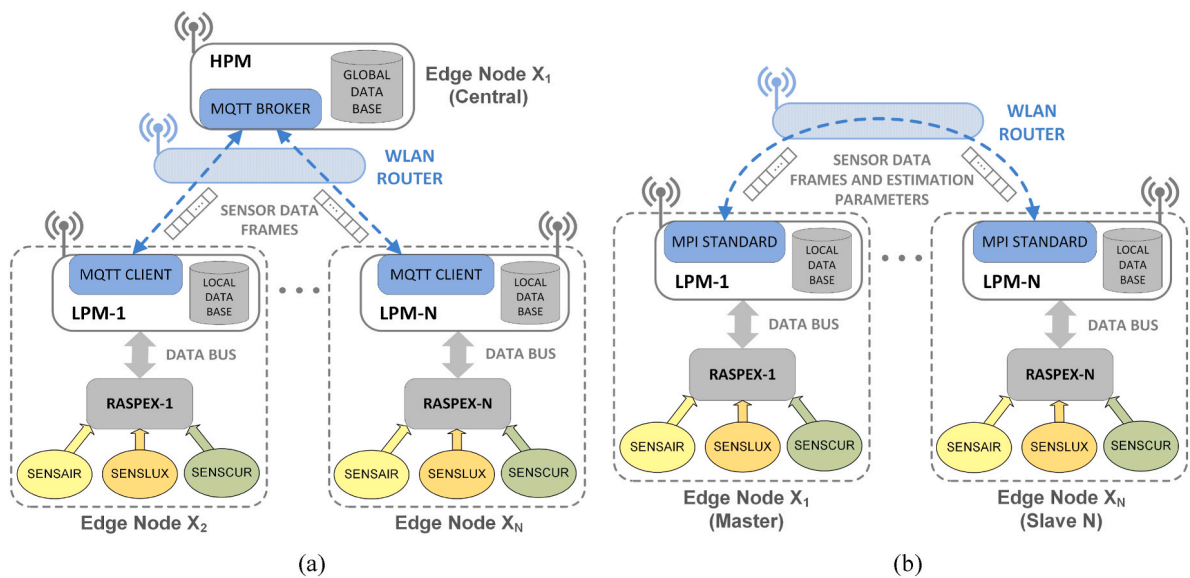

**Fig. 1.** Schematic layouts of the proposed IoT architectures: (a) centralised and (b) distributed parallel.

*2.2.1. SENSAIR module*

This module was implemented with the Sensirion SVM30 board to obtain data from the environmental parameters, such as temperature, RH, and some different gases such as $CO_2$ and VOC, which were measured and processed with the experimental setup. It includes automated calibrated air quality output signals with baseline compensation as well as humidity compensation of metal oxide (MOX) gas sensors. This module was connected to the processing module via the Inter-Integrated Circuit ($I^2C$) bus. For instance, the temperature and RH equations to obtain the analog values once acquired and conditioned by the SVM30 board are the following:

$$Temp = -45.7 + 175.7 \bullet \frac{S_t}{2^{16}} \quad (1)$$

$$RH = \left(103.7 - 3.2 \bullet \frac{S_t}{2^{16}}\right) \bullet \frac{S_{rh}}{2^{16}} \quad (2)$$

where $S_t$ and $S_{rh}$ are the values of temperature and RH in raw format, respectively.

*2.2.2. SENSLUX module*

This module provided illuminance data and was implemented with the Luna Optoelectronics NSL-19M51 sensor. Its parameters are listed in Table A.2 of Appendix A and the measurements were only used to distinguish the main time slots of a day, so five digital scales were sufficient after conversion with the 10-bit ADC (range 0–1023). These scales were: Very Low (VL) for night at 0–204, Low (L) for sunset at 205–410, Medium (M) for sunrise at 411–616, High (H) for afternoon at 617–822, and Very High (VH) for noon at 823–1023.

*2.2.3. SENSCUR module*

As discussed above, this module was implemented using the Mikroe MIKROE-1396 board which mainly included an INA196 current shunt monitor and a MAX6106 voltage reference. This board received current from either LPM or HPM processing modules in the range 100–2048 mA, and was then converted through the INA196 to a voltage in the range of 0.1–2.048 V to be connected to a 10-bit ADC channel on the RASPEX board. The RASPEX board contained a MCP3008 ADC and was attached to the processing module via the Serial Peripheral Interface (SPI) bus. To improve measurement resolution, the minimum current was set to 400 mA since it was verified that the processing modules did not consume less in standby mode, which is the operating mode between periodic acquisitions of sensor data without data handling. The output of the 10-bit ADC was a value $x$ in the range 0–1023 and sent to the processing module via SPI to obtain a differential analog voltage (in Volts) with the following linear equation:

$$V_{an} = 1.6 \bullet 10^{-3} \, x + 0.4 \quad (3)$$

The equivalent current (in Amperes) was obtained according to the selected configuration of the INA196, as indicated in the equation below:

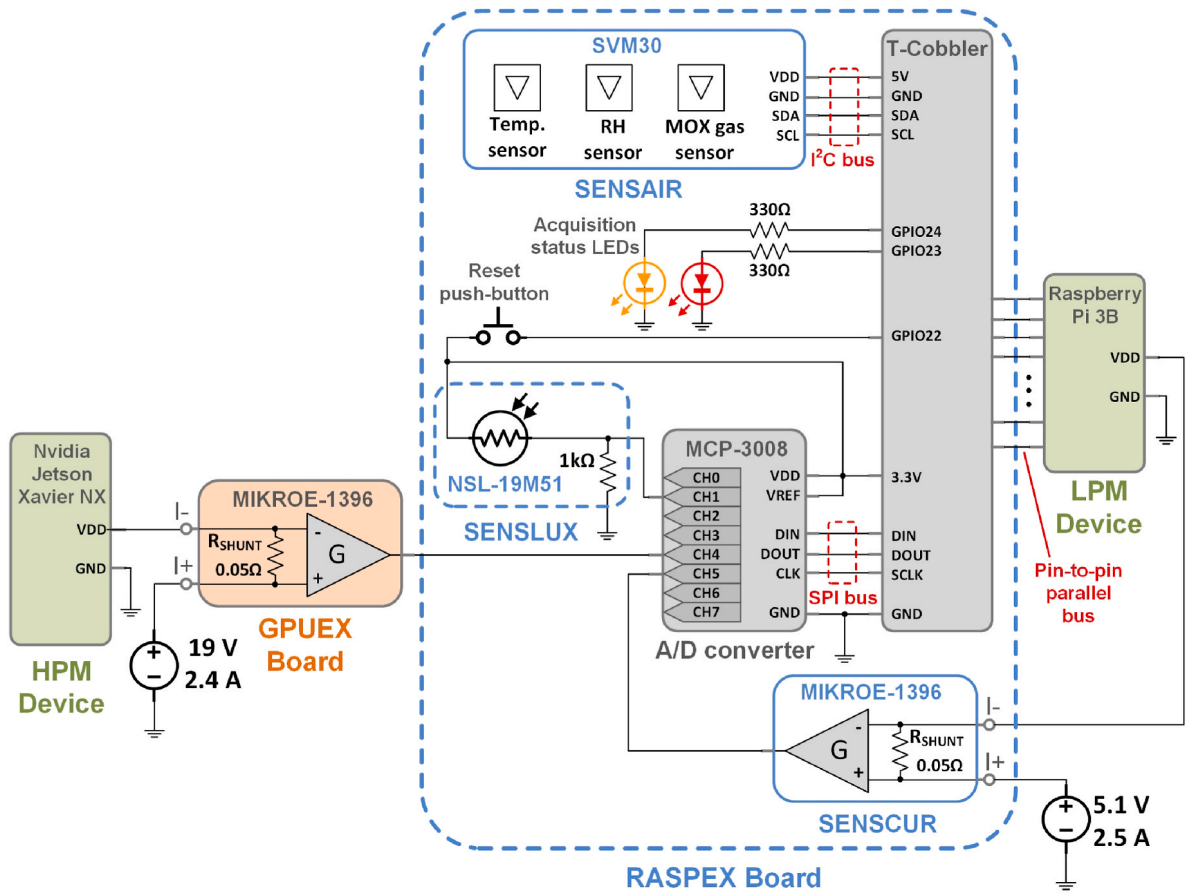

**Fig. 3.** Electronic design of the RASPEX board and connection to GPUEX board and LPM/HPM processing devices.

$$I_s = \frac{V_{an} \bullet R_1}{R_S \bullet R_L} \tag{4}$$

where $R_1$ is an internal resistor with a fixed value of 5 kΩ, $R_L$ is the load resistor before the output buffer amplifier with a value of 100 kΩ to get a gain of 20 V/V, and $R_S$ is the shunt resistor of value 0.05 Ω to obtain a current in the range 400–2048 mA.

### 2.3. Sensor data frames

Sensor data was exchanged between processing modules using an ad-hoc format to encapsulate the RASPEX identifier, acquisition timestamp, sensor types, and measurement values, among other fields. Fig. 4 shows the structure of the communications data frame, with length *N*, used in both architectures. A complete description of the data frame fields is included in Table A.3 of Appendix A. The data frame was designed to support more sensors for environmental measurements than those used in the presented paper, such as noise or motion, to allow for scalability with different edge nodes. As discussed above, the encapsulated information in a data frame was sent via MQTT in the centralised architecture or via MPI in the parallel distributed one. The destination module unpacked the received frame and stored the data in timestamp order in its local database.

Fig. A.1-a of Appendix A shows the specific data frame format used in the presented paper with a size of 49 bytes, which includes all fields for environmental parameters and current consumption of LPM and HPM modules, collected in each acquisition period. This data frame was designed to minimise communication overhead between devices, and to optimise the efficiency of data transmission with MQTT and MPI, all data measured in one period from each sensor was included in the same data frame to use the remaining time between consecutive acquisitions for processing tasks of the ML models (Alam et al., 2017). The measured data with the RASPEX boards was encapsulated using the IEEE 754 format (4 bytes) in the "sensor measure" field of the data frame, as shown in Fig. A.1-b of Appendix A with an example of a temperature measurement. An IEEE 754 formatted measurement value is stored in binary format and then converted to decimal format in groups of eight bits to be stored at byte-level in a one-dimensional vector of unsigned char data type. Once all the data was encapsulated, it was sent using the MQTT protocol with *publish/subscribe* functions or with the MPI standard using the *MPI_Send* primitive with the entire data frame in the "buffer" parameter (Quinn, 2003; Wilkinson and Allen, 2004).

### 2.4. Software components

The implementation of the software components was carried out with the C and Python programming languages using a wide range of external commercial libraries, as shown in Table A.4 of Appendix A. The C language was used for low-level programming to implement the data acquisition (WiringPi and Svm30 libraries), the use of MQTT protocol (Mosquitto distribution), and storage of measured data in LPM local databases (Mysql library). However, the Python language was used to store data in the HPM global database (MysqlDB library), the use of MPI standard (MPI4py), and to develop the ML models using external libraries, such as Tensorflow, Pandas, Numpy and Matplotlib.

### 2.5. Database management

As already stated, MariaDB was used for local and global databases,

as it is the open source solution using the MySQL engine, which can be used on aarch64 (ARM64) processor architectures of LPMs and HPM with the libraries listed in Table A.4 of Appendix A. Prior to its use, it was compared with SQLite to identify which ones would be a better fit for the developed work. Although, SQLite is a much faster database technology and its simplicity of implementation is aligned with the low resource requirements of the presented article, MySQL offers more security and scalability than SQLite. Therefore, it was selected since scalability is one of the main objectives of the presented work. In addition, other options that could fit into this project were evaluated due to their consistency and scalability, such as NewSQL (Muniswamaiah et al., 2020). Fig. A.2 of Appendix A shows a view of the database with the measurements of all environmental and current sensors in each acquisition period (see "date_time" timestamp).

## 3. Experimental methods for estimating environmental parameters

This section describes the proposed methods to estimate environmental parameters based on centralised and distributed parallel IoT architectures. Firstly, evaluation guidelines and the general estimation methodology are introduced to obtain a broad perspective of the fundamentals of both methods. Next, MLP-ANN models are considered as the ML algorithms implemented to evaluate both methods, and specifically the details of their training process are analysed. Subsequently, the complete process structured in phases that were implemented for both methods is described in detail, from data acquisition to obtaining local and global forecasts, through the training and testing processes.

### 3.1. Evaluation guidelines of estimation methods

To evaluate the performance of the proposed methods, not only the success of the training was considered, but also the criteria for a correct assessment of the estimation results with respect to the expected values. It was carried out with fractional parts using random ordering of ground truth data for training and test datasets, and the selection of a sufficiently simple and flexible learning methodology. Likewise, as the statistical criteria played a relevant role in the evaluation, on the base of the confusion matrix the F-score and the Accuracy were used as classification metrics. The confusion matrix shown in Table 1 was used with the indicated labelling criteria of estimated and expected outputs for a specific experimental test.

The F-score (or $F_1$-score) is the harmonic mean of precision and recall, and was used as a combined evaluation metric to assess average rates rather than the Accuracy. However, the Accuracy also provided an additional ponderation as evaluates the ratio of true cases (estimated correctly) and the total number of examples. Their formulas are reminded next:

$$F - score = 2 \bullet \frac{P \bullet R}{P + R} \tag{5}$$

$$Accuracy = \frac{\#True\ positive + \#True\ negative}{\#Total\ samples} \tag{6}$$

where *P* is the Precision as the ratio of true positives and the number of predicted positive cases (true and false positives), and *R* is the Recall as the ratio of true positives and the number of actual positive cases (true

**Table 1**
Confusion matrix with labelling criteria of expected and estimated outputs.

| Confusion matrix | | Expected outputs | |
|---|---|---|---|
| | | $y^{(i)}_{label} = 1$ | $y^{(i)}_{label} = 0$ |
| Estimated outputs | $\hat{y}^{(i)}_{label} = 1$ | #True positive | #False positive |
| | $\hat{y}^{(i)}_{label} = 0$ | #False negative | #True negative |

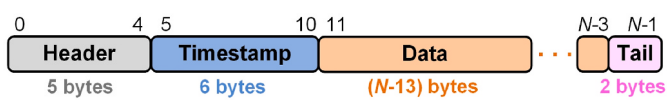

**Fig. 4.** Generic format of the data frame to encapsulate the sensor data.

positives and false negatives) as follows:

$$Precision = \frac{\#True\ positive}{\#True\ positive + \#False\ positive} \quad (7)$$

$$Recall = \frac{\#True\ positive}{\#True\ positive + \#False\ negative} \quad (8)$$

The training performance evaluation was performed using the Training Accuracy as indicated below:

$$TA = \frac{\sum_{i=1}^{m}(|\widehat{y}_i - y_i| < TOL)}{m} \quad (9)$$

where $m$ is the number of training examples, $y^{(i)}$ is the expected output to train the model, $\widehat{y}^{(i)}$ is the estimated output from the model, and $TOL$ is the accepted tolerance for considering an estimation as valid, such as 0.01 °C for temperature forecast.

In addition to the classification metrics indicated above, in the article the RMSE has been considered in relation to the IEQ as it is the most used metric in building simulation analysis and the one recommended by ASHRAE (ASHRAE, 2023; Martínez et al., 2020). In addition, the MAE is also considered as a common performance metric for ML applied to IEQ, as it quantifies the difference between the matched estimated and target values over a large set of examples, and was also intended to be minimised with an optimal set of ANN parameters $\Theta^{(j)}$ as network weights. This is a typical technique of measuring a model errors of air quality with the prediction of quantitative data and is a sort of separation between the estimated values and the actual values (Almalawi et al., 2022).

$$RMSE = \sqrt{\frac{1}{n}\sum_{i=1}^{n}(\widehat{y}_i - y_i)^2} \quad (10)$$

$$MAE = \frac{1}{n}\sum_{i=1}^{n}|\widehat{y}_i - y_i| \quad (11)$$

where $n$ is the total number of examples.

Only for the distributed parallel architecture, parallel processing metrics were also considered to assess the processing performance:

$$Speedup = \frac{T_s}{T_p} \quad (12)$$

$$Efficiency = \frac{Speedup}{Number\ of\ Processors} \quad (13)$$

where $T_s$ and $T_p$ are the sequential and parallel processing times, respectively.

### 3.2. General estimation methodology

The methodology applied for the conditioning of the dataset with environmental inputs and expected outputs, the training and testing of the ML models, the final debugging, and obtaining the estimation performance metrics consisted of seven steps, as shown in Fig. 5.

In the step 1, environmental parameters were measured and acquired periodically (acquisition period $T_{ac}$) by each edge node and stored in the local databases of LPM and HPM devices, which represented the input data $x^{(i)}$ of the dataset.

In the step 2, it was stablished a criterion to map the sensor data (model inputs) with the expected output $y^{(i)}$ and label that output for each example $(x^{(i)}, y^{(i)})$ to allow models to learn. In general, for any type of environmental parameter acquired with the presented architectures, this criterion was as follows:

$$y^{(i)} = \begin{cases} Computed\ Value & if\ y^{(i)} \in Range \rightarrow y^{(i)}_{label} = 1\ (Valid, True) \\ Fixed\ Value & if\ y^{(i)} \notin Range \rightarrow y^{(i)}_{label} = 0\ (Invalid, False) \end{cases} \quad (14)$$

where the Computed Value is obtained from an equation that relates the model inputs, the Fixed Value is the minimum or the maximum limit of the range to set a value that shall not be exceeded, and $y^{(i)}_{label}$ is the label assigned to the training example $i$.

In the step 3, after the expected outputs were assigned and each example in the dataset was labelled, all the examples of ground truth data were randomly reordered for symmetry breaking. Then, the dataset was divided into a 70% fraction for training and a 30% fraction for testing/validation (Shah et al., 2020; Wei et al., 2019) to evaluate the generalizability property of the models and obtain an independent measure of performance after training. Since the acquired data could contain noise, missing values, etc., it was pre-processed prior to the training step to develop robust estimation models and for which some techniques were applied, such as outlier detection and data normalisation (Zhang and Woo, 2020).

In the step 4, the model was trained and the training errors were computed using the considered cost function until reaching at least 0.5–0.75 TA. This accuracy value was not set too high to reduce processing time and power consumption when using small training sets. In addition, in this step some learning curves were obtained to check that the training loss decreased to a point of stability, which provided an idea of how well the model is learning to avoid wasting time collecting more training data that is useful and identifying underfitting in the model. For training, one of the most relevant points was to minimise the cost function by finding the proper weights of the links of the nodes in the ANN to ensure a good generalization. In order to generate better weights in each training iteration (epoch), an optimization algorithm was selected to compute the cost function gradient $\frac{dJ(\theta)}{d\theta_j}$ to allow each input feature weight to be improved at each iteration as follows:

$$\theta_j^{t+1} = \theta_j^t - \alpha \bullet \frac{dJ(\theta)}{d\theta_j^t} \quad (j = 0, 1, \ldots n) \tag{15}$$

where $\theta_j^t$ is the weight for the input feature $j$ in the iteration $t$, $n$ is the number input features, $J(\theta)$ is the cost function, and $\alpha$ is the learning rate.

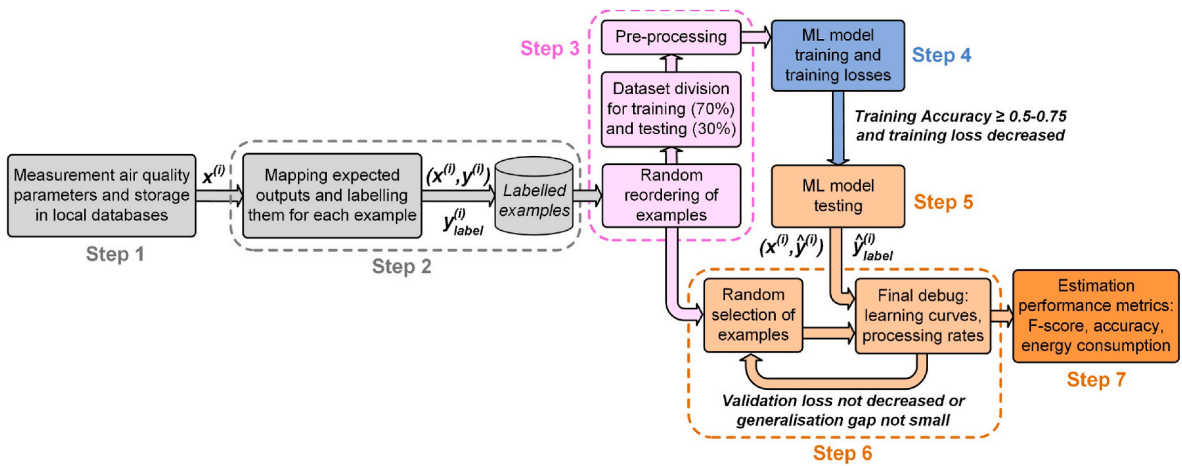

**Fig. 5.** Process of the general estimation methodology.

In the step 5, the final model was evaluated using the test dataset one the TA limit was reached. The model estimations $\widehat{y}^{(i)}$ were obtained from sensor data used only for testing $x_{test}^{(i)}$. In addition, it was necessary to consider a criterion to label an output estimation as correct or not with respect to the expected output $y^{(i)}$, for which an adequate tolerance (*TOL*) was established according to the type of air quality parameter measured and the resolution of the sensor used. In general, for any type of air quality parameter acquired with the presented architectures, this criterion was as follows:

$$\left| y^{(i)} - \widehat{y}^{(i)} \right| = \begin{cases} Value \leq TOL & \to \widehat{y}_{label}^{(i)} = 1 \; (Valid, True) \\ Value > TOL & \to \widehat{y}_{label}^{(i)} = 0 \; (Invalid, False) \end{cases} \tag{16}$$

where $\widehat{y}_{label}^{(i)}$ is the label assigned to the estimated output of the test example $i$.

In the step 6, several learning curves with randomly selected examples were plotted to check that the validation loss decreased to a point of stability with a small generalisation gap (gap between the train and validation loss learning curves). It provided insight into how well the model is generalizing and, if it exists, showed the dynamics of overfitting, which was especially important for small training sets like in the presented paper. For this task, several examples were randomly selected from the training and test sets, then the model parameters $\theta$ were learned using the randomly chosen training set, and these parameters were evaluated on the randomly chosen training and test sets. This process was iteratively repeated several times until a suitable rapid convergence of loss (error) to zero was reached in the learning curves, and the power consumption of the processing considering CPU usage, RAM memory, etc. was appropriate.

Finally, in the step 7, the final metrics based on the confusion matrix, such as F-score and Accuracy, were obtained to evaluate the estimation performance and they were contrasted with power consumption. It was used to draw conclusions about the most appropriate ML models regarding the estimation accuracy versus efficiency in power consumption and resource usage, which were key factors in the proposed edge computing-based IoT architectures.

The following subsections explain the extension of the general estimation methodology to deploy ML models using methods based on both proposed architectures. Some Federated Learning principles resemble to some extent some step in implemented estimation methods, such as the Federated Averaging approach that allows combining local and global model averaging, and transmitting training model data to a central node periodically at strategic intervals instead of continuously (Wang et al., 2020).

### 3.3. Training process of MLP-ANN models

After the step 3 in the general methodology, the training set was obtained to develop an MLP-ANN with a specific topology. Fig. 6 shows the detailed process and libraries used to train a selected MLP-ANN according to the step 4 of Fig. 5, whose topology was developed using the library Keras over TensorFlow (TF) with *TF.keras.layers.dense*. In this process, the optimization algorithm (optimizer) selected to readjust the network weights was Adam that combines the properties of the optimisers AdaGrad (Adaptive Gradient Algorithm) and RMSProp (Root Mean Square Propagation), the loss type was the Mean Squared Error (MSE) type to verify the error between the estimations and the expected outputs to obtain the TA, and the learning rate was 0.01 to get fast convergence of the optimization algorithm after trying other values. The model was also fitted with this library using the estimated and expected values, and selecting the number of training epochs as a parameter, which allowed obtaining data to plot some learning curves using the Matplotlib library. The training of the MLP-ANN was performed once $m$ examples $(x^{(i)}, y^{(i)})$ were acquired, where $i = 1$ to $m$, $x = x^{(i)}$ is the input vector of sensor data, and $y = y^{(i)}$ is the output vector of expected values. The weights of each ANN node were randomly initialised and the sensor data $x$ was used as ANN input data ($a^{(1)} = x$). For simplicity, if a three-layer MLP-ANN is considered, in the forward propagation phase of the algorithm, the vectors of the remaining neurons $a^{(j)}$ were calculated, where $j = 2$ (hidden layer) and $j = 3$ (output layer). The ANN hypothesis output vector $h_\Theta(x)$ was calculated as expressed below:

$$h_\Theta(x) = a^{(3)} = g_2\left(\Theta^{(2)} \bullet a^{(2)}\right) \tag{17}$$

$$a^{(2)} = g_1\left(\Theta^{(1)} \bullet a^{(1)}\right) \tag{18}$$

where $g_1(x) = -1 + 2/(1 + exp(-2 \bullet x))$ is the tan-sigmoid activation function for the hidden layers to establish nonlinear relations among layer inputs and outputs with a limited range to $\pm 1$, $g_2$ is the activation function for the output layer, $\Theta^{(1)}$ is the weight matrix to control the mapping of the input layer to the hidden layer, and $\Theta^{(2)}$ is the weight matrix to control the mapping of the hidden layer to the output layer. It was admitted that an MLP-ANN with tan-sigmoid hidden neurons can adapt quite well to multidimensional mapping problems when fed with consistent data and the hidden layer has enough neurons.

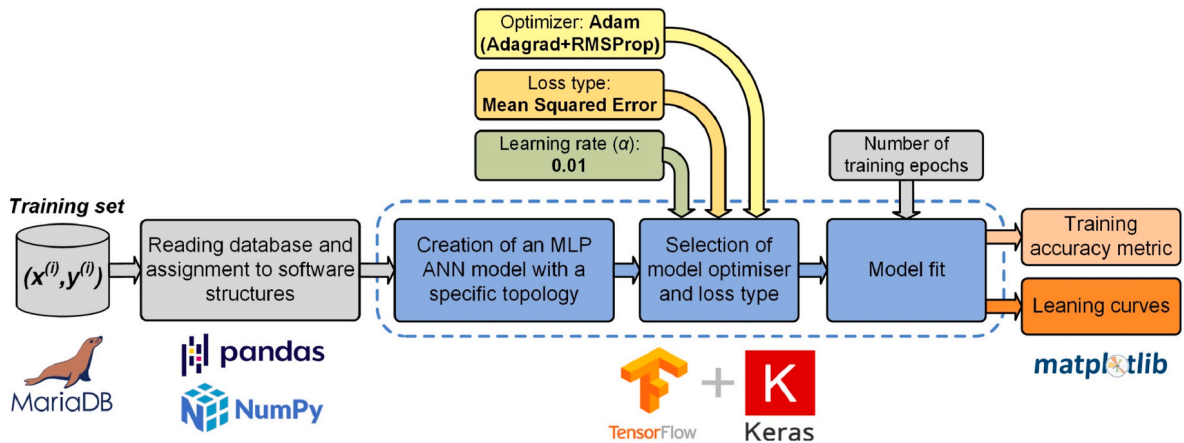

**Fig. 6.** Training process of MLP-ANN models.

According to a statistical learning approach, more hidden layers mean more complex estimation models, which improves the ability to interpret data patterns. However, if the number of hidden layers is increased excessively, gradient vanishing or gradient exploding effects could occur during training (Hastie et al., 2009). These problems could lead to overfitting and a prolongation of the training process. To avoid overfitting and shorten the training time, the current work established a maximum of two and three hidden layers to search for the best prediction performance of initial models, so four-layer and five-layer MLP-ANN topologies were implemented. The general topology of MLP-ANN models is shown in Fig. 7, which is composed of an input layer with a number of neurons depending on the number of edge IoT nodes deployed in the architecture, two or more hidden layers with a variable number of neurons depending of the best fit obtained in the design, and an output layer with only one neuron that provides the estimated output. At the input layer, instantaneous and differential inputs were considered in the topologies, but at least the instantaneous were necessary and differential were optional to implement models more complex that needed additional information about input changes over time. Based on the general MLP-ANN topology, the case study for temperature estimation was performed with 3-2-2-1, 3-6-6-2 and 3-50-50-1 topologies in the centralised architecture, as shown in Fig. B.1 of Appendix B, and a 3-10-10-10-1 topology in the parallel distributed architecture.

### 3.4. Centralised estimation method

In this method, the complete process of which is shown in Fig. 8, all edge nodes periodically acquired data from the SENSAIR, SENSLUX and SENSCUR sensor modules with an acquisition period $T_{ac}$. To obtain estimation metrics and environmental global forecast, five phases were implemented in this method as follows (for reference only, some phases are analysed in detail in the subsections and figures above).

- *1st phase:* all the nodes acquired and locally stored a great amount of data before sending it to the central node ($X_1$) without performing any additional processing operations.
- *2nd phase:* the data was encapsulated according to the data frame format shown in Fig. 4 and sent to the central node by MQTT to perform data fusion, map expected outputs and label them for each example (see Fig. 5-step 2), and split the data into training and test datasets.
- *3rd phase:* the model training (see Fig. 5-step 4 and Fig. 6) was carried out in the central node until a TA of at least 0.85 was obtained. While the training was in progress, all nodes continued to acquire data and store it in local databases. When the training process was finished and the TA did not reach at least 0.85, the central node received more data from other nodes by MQTT as in the 2nd phase to continue training. In this phase it was important to prioritise training with few data and apply, if necessary, TinyML techniques such as quantisation and pruning (Raha et al., 2021; Sanchez-Iborra and Skarmeta, 2020) to try not to have overfitting (high variance) in the model and improve the generalization property.
- *4th phase:* if the training was successful (TA $\geq$ 0.85), then the model weights ($\theta$) were obtained and the testing process (see Fig. 5-step 5) started to continue with the final debug and obtain the estimation metrics (see Fig. 5-steps 6 and 7). Specifically, the F-score metric was used as a simple criterion to evaluate the estimation performance, considering a value of 0.9 as passing threshold to consider an adequate model to obtain the global forecast.
- *5th phase:* if the testing was unsuccessful (F-score < 0.9) then the central node received more data from other nodes as in the 2nd phase. When the testing was successful, the first global forecast was obtained and then all nodes periodically sent new sensor data to obtain near real-time forecasts. In that period of 10 $T_{ac}$, the training process and subsequent testing were performed again, taking the previous weights as a base and using in each period a dataset of 10 $N$ new examples to improve learning ($N$ is the number of edge nodes including the central node). In this way it is possible to adjust the model with the new data received to obtain forecasts adapted to changes in environmental parameters.

### 3.5. Distributed parallel estimation method

In this method, the complete process of which is shown in Fig. 9, and similar to the centralised one, all the edge nodes periodically acquired data from the sensor modules with an acquisition period $T_{ac}$. However, local predictions are now obtained at each edge node, and estimation metrics and environmental global forecast are only obtained at the master node. The method was implemented in seven phases as follows, slightly more complex than the centralised method (for reference only, some phases are explained in detail in the subsections and figures above).

- *1st phase:* all the nodes acquired and locally stored a great amount of data for further training and testing. This data is already stored in local databases as examples mapped to expected outputs and labelled (see Fig. 5-step 2). In this phase, the data was not sent to the master node ($X_1$) but was used to perform processing operations to learn ML models locally.
- *2nd phase:* the data was split into training and test datasets at each edge node (master and slaves), and model training (see Fig. 5-step 4 and Fig. 6) was carried out in parallel at each node until a TA of at least 0.85 was reached. While the training was in progress, all nodes continued to acquire data and store it in local databases. When the training process was finished and the TA did not reach at least 0.85, the nodes read more data from their local databases to continue training. Again, in this phase it was important to prioritise training with few data and apply, if necessary, TinyML techniques such as quantisation and pruning (Raha et al., 2021; Sanchez-Iborra and Skarmeta, 2020) to avoid overfitting and improve the generalization property. Compared to the centralised method, now the dataset splitting, training, and TA threshold verification were performed in parallel at each edge node, allowing nodes with less computing capacity to be used instead of having to a much more powerful central node.

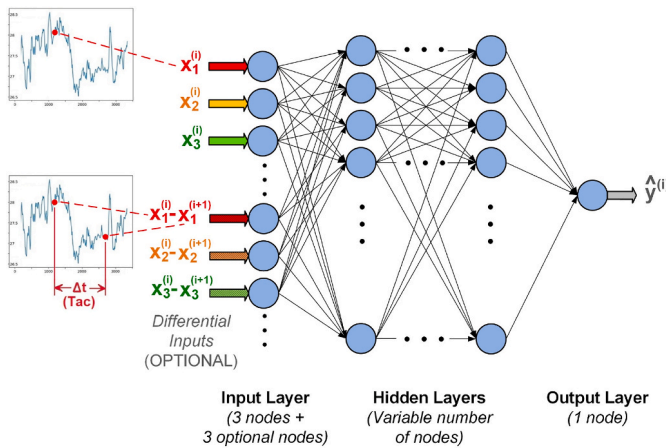

**Fig. 7.** General topology of the implemented MLP-ANN models.

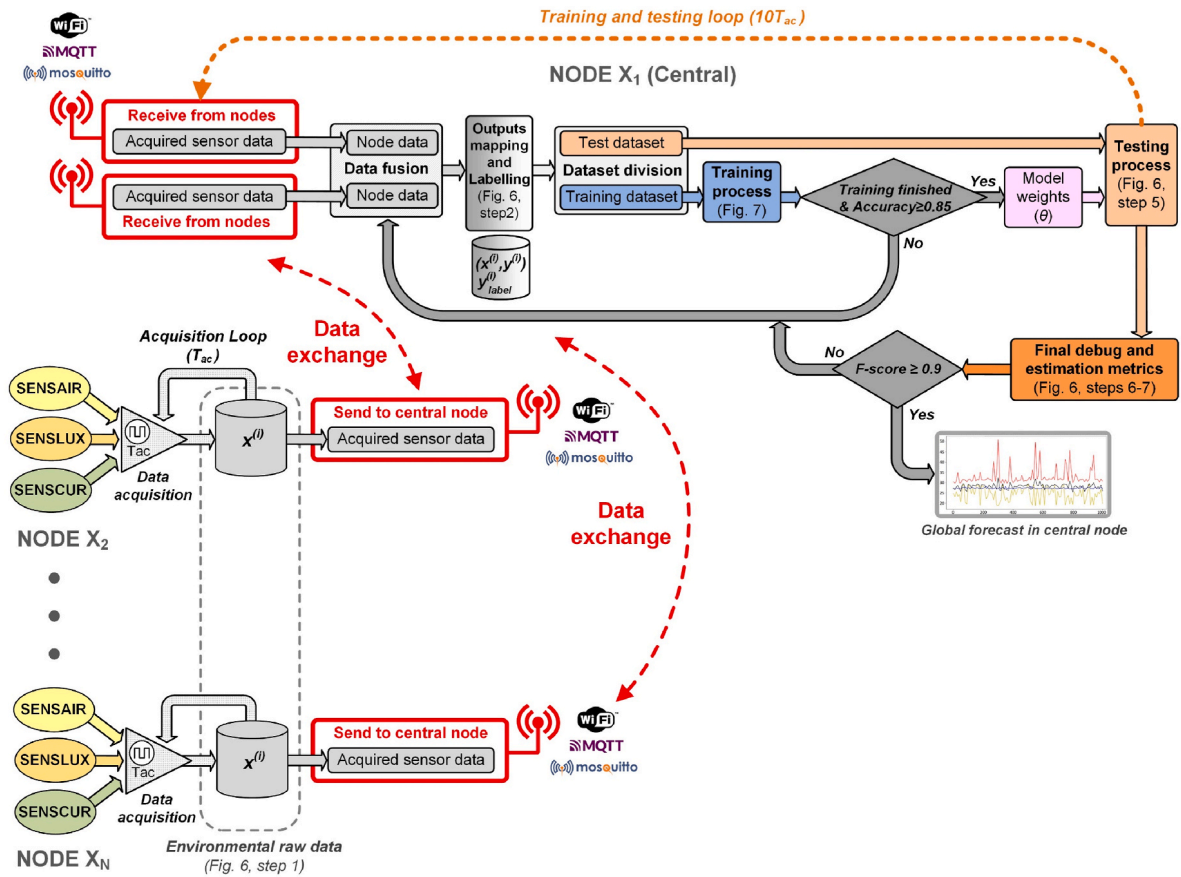

**Fig. 8.** Complete process of the centralised estimation method.

- *3rd phase:* if the training was successful (TA ≥ 0.85), then the model weights ($\theta$) and the testing dataset obtained in slave nodes were encapsulated according to the data frame format shown in Fig. 4 and sent to the master node by MPI. In addition, local forecasts were obtained at edge nodes as preliminary results using the model learned locally.
- *4th phase:* the master node received by MPI local test datasets and model weights of the slave nodes, and together with the weights obtained in the master, the average per weight was calculated ($\overline{\theta}$) to obtain the weight matrix ($\overline{\Theta}$) for the "aggregated" model that served to start the testing process.
- *5th phase:* the testing process (see Fig. 5-step 5) started to continue with the final debug and obtain the estimation metrics (see Fig. 5- steps 6 and 7). Again, the F-score metric was used to evaluate the estimation performance, considering a value of 0.9 as passing threshold to consider an adequate model to obtain the global forecast.
- *6th phase:* if the testing was unsuccessful (F-score < 0.9) then the process started again in the 2nd phase, that is, the master node read more data from its local database, performed the training again, and obtained the model weights (taking the previous weights as a base) that were averaged with slaves' model weights received by MPI. When the testing was successful, the master node sent the averaged model weights to slaves and obtained the first global forecast.
- *7th phase:* similar to the centralised method, in a period of 10 $T_{ac}$ the training process is performed again in all edge nodes, taking the averaged weights as a base and providing in this period a dataset of ten new examples. In this way it is possible to adjust the models with the new data to obtain local forecasts and global forecast in near real-time adapted to changes in environmental parameters.

## 4. Experimental results and discussion

The main objective was to assess that the estimation of environmental parameters was carried out correctly with the proposed methodology using both IoT architectures and the ML algorithms without influencing deviations or outliers, and relate it to the power consumption of the processing. The experimental setup is shown in Figs. 1 and 2, in which the processing modules with the characteristics listed in Table A.1 and the sensor modules with the parameters shown in Table A.2 were used, both tables included in Appendix A. The models were evaluated on datasets obtained from multiple peak, step, and transient test cases for each environmental parameter, although the performance analysis was focused on temperature test scenarios. The performance was analysed with two experiments focused on temperature and related to power consumption data, and, in auxiliary way, using illuminance variations. It should be noted that the proposed methodology and some of the results can be extrapolated to other environmental parameters that were measured in the experiments, but whose analysis is not included for brevity.

### 4.1. Design of the experiments and data acquisition

The experiments were carried out in an open-plan indoor environment with a dimension of approximately 41.7 m² and natural light windows to achieve representative distributions of air quality parameters. The three IoT edge nodes (*X*1, *X*2, and *X*3) were placed in the room to have enough distance between them and use available ventilation resources, such as two wall-mounted HVAC (Heating, Ventilation, and Air Conditioning) equipment and a window to place a node in the outside, as shown in the floor plan of Fig. 10. The location photos of edge nodes *X*2 and *X*3 in the room are shown in Fig. B.2-a and Fig. B.2-b of

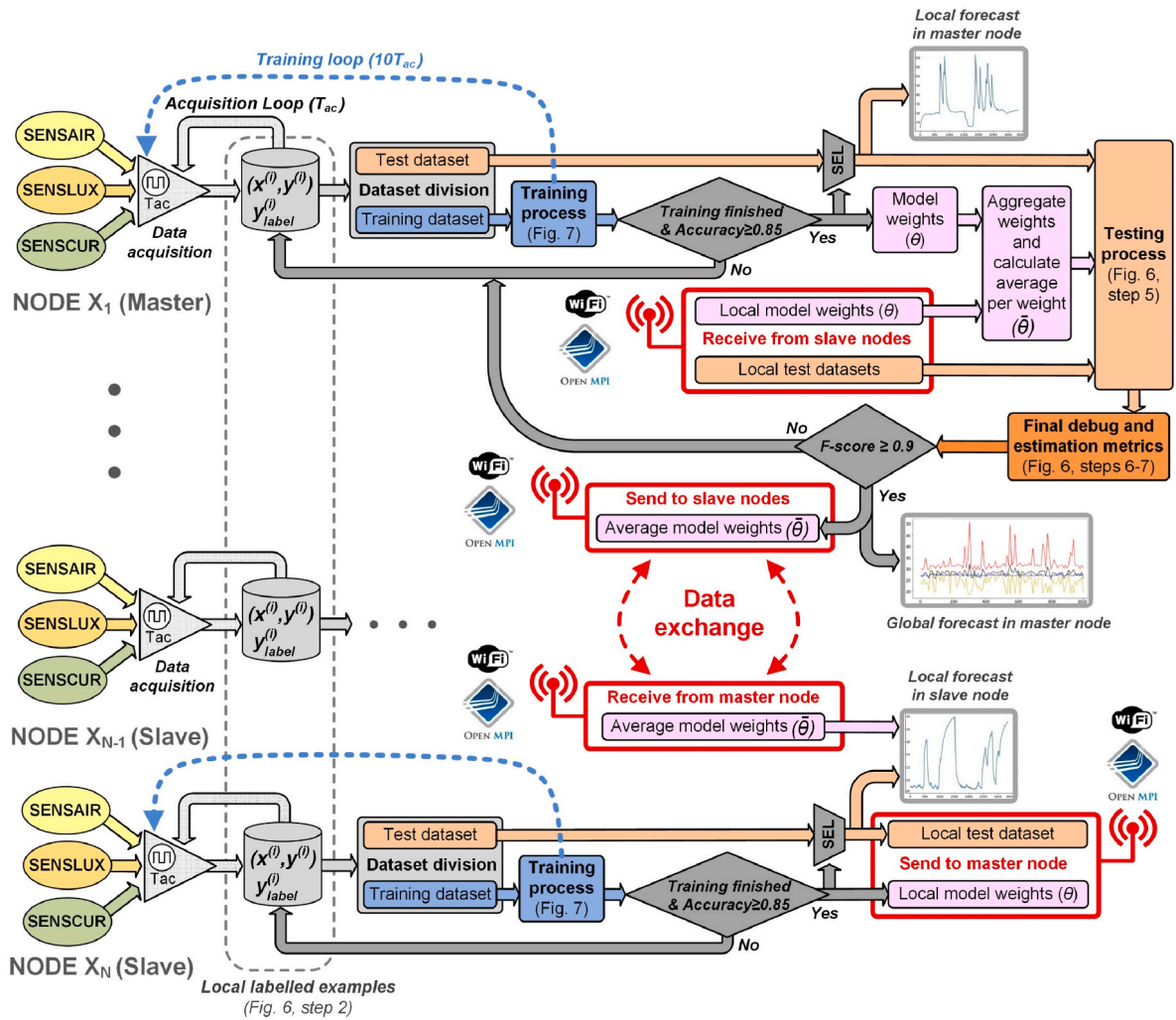

**Fig. 9.** Complete process of the distributed parallel estimation method.

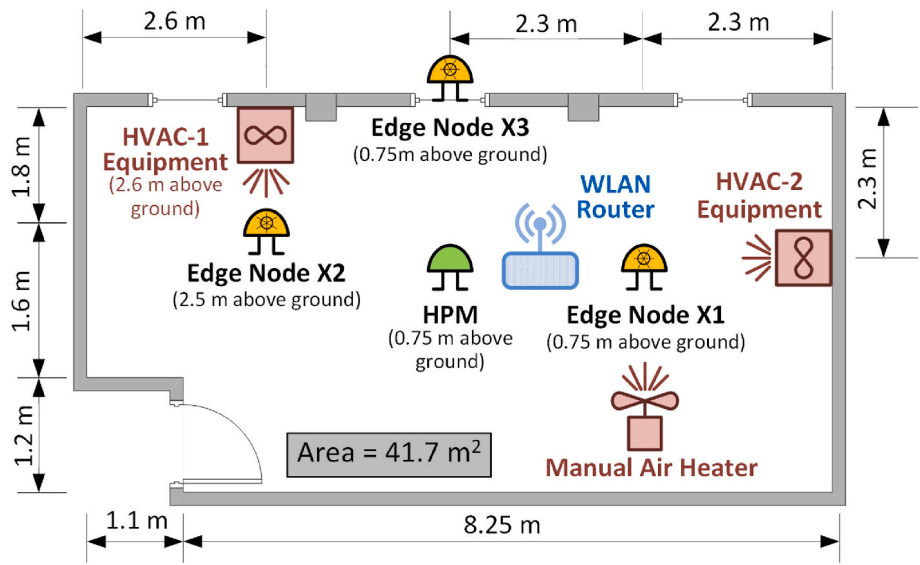

**Fig. 10.** Indoor floor plan for experimental tests with the prototype modules distributed in the room.

Appendix B, respectively.

According to section 2, each edge node collected six sensor measurements per acquisition period ($T_{ac}$) from each sensor module: SENSAIR module with four parameters (temperature, RH, $CO_2$, and VOC), SENSLUX module with one parameter (illuminance), and SENSCUR module with one parameter (DC current) for each LPM and HPM.

Therefore, in each $T_{ac}$ six measurements were obtained for each LPM edge node and one measurement per HPM edge node, obtaining a total of 19 measurements per sampling period. In order to have significant changes in air quality measurements, it is common to take samples in periods of 45–60 min and 24 h per day for 12–15 days. To reduce training and testing times, $T_{ac}$ was 10 s and the temperature profiles of two experiments were performed for almost an hour, obtaining a dataset with measurements equivalent to taking sensor samples for 15 days and $T_{ac}$ of 60 min. This sampling rate was set to the minimum level allowed because reducing the frequency of the samples taken could optimise the power consumption and memory storage, which could be especially useful in the case of battery-powered systems. As described in Table 2, the experimental training, validation and testing of the models were accomplished on 6460 measurements of parameters with temperature profiles with a variety of shapes, such as sudden spikes, square waveforms, and sawtooth waveforms between cold (18 °C) and heat (50–60 °C) over normal temperature (27.5 °C) or cold temperature baselines for summer in Spain.

The Experiment 1 was the most important and was mainly focused on the validation of the centralised architecture, but the distributed parallel architecture was also partially validated with the same test profile as the centralised one to be able to compare the performance of both. Due to the operational complexity to validate the ANN model in the distributed parallel architecture, the Experiment 2 was only run in this architecture with a different test profile. The datasets of each experiment were different, more normal in Experiment 1 and more extreme in Experiment 2. The Experiment 1 consisted of three temperature profiles with indoor sudden cold-hot or hot spikes from the normal and cold baselines, combined with a stable outdoor temperature typical of summer in Spain. The Experiment 2 consisted of three temperature profiles somewhat more extreme than Experiment 1, including not only cold-hot spikes, but also high-frequency sawtooth profiles both indoors and outdoors, and two bursts of cold-hot sawtooth in the outside temperature. Next, the configuration of edge nodes $X1$, $X2$ and $X3$ in both experiments are explained in detail below, which in the in the distributed parallel architecture were configured $X1$ as the master node, $X2$ as the slave node 1, and $X3$ as the slave node 2. Common to both, the hot spikes were obtained using a manual air heater, a wall-mounted HVAC and ice cubes with a manual fan were used to generate cold wind over the entire nodes.

In the Experiment 1, the $X1$ node was selected as the most unstable data source due to it received temperature cold and hot spikes, the $X2$ node was only subjected to temperature drops using the wall-mounted HVAC, and the $X3$ node was placed on the window of the room and was dedicated to obtaining the reference outdoor temperature. As depicted in Fig. 11, the $X1$ node obtained data without being exposed to modifications during the first 780 s. From this moment on, the manual air heater was used to heat the node in spikes and let it cool down to the instant 1200 s. Up to the instant 1900 s it was not undergone changes, it was then that the ice cubes were used to lower the temperature until the instant 2100 s. Thereafter, the node was again subjected to hot spikes until 2800 s, and finally no changes were made until the end of the experiment at 3600 s. Regarding the $X2$ node, it was hung on the ceiling of the room so that it was only subjected to temperature drops from a normal temperature baseline using the wall-mounted HVAC. The $X2$ temperature profile started with the HVAC turned on, receiving cold air until the instant 600 s. Up to the instant 780 s it was not subjected to any stimulus and after that the HVAC was activated again until the instant 1200 s. The next part of the test profile up to the instant 1900 s passed without cold air. From that instant to 3000 s the HVAC was turned on, and since then it was turned off and on every 120 s until the end of the experiment. The actual temperature data acquired by each edge node and the fusion of these data to represent the full profile of the Experiment 1 are shown in Fig. 12.

In addition, the illuminance data acquired by each edge node was used to estimate the time slot of the day in order to have an auxiliary reference for a proper analysis of the temperature estimation. As commented in section 2, the illuminance sensor was connected to an ADC channel and it returned a digital value between 0 and 1023, depending on the amount of light that falls on it. Because this measurement was not very significant in the Experiment 1, the sensor was calibrated to identify the time slot of the day approximately so it was considered the value of luminosity on a scale from 1 to 5 according to Table 3. Under these considerations for illuminance, as shown in Fig. 13, the $X1$ node light sensor was used as an unstable data source due to receiving steps of very dark and dark to high illuminance (night to afternoon), the $X2$ node sensor was forced to have darkness, so its value remained stable in very dark illuminance (night), and the $X3$ node sensor was forced to have a step of very high to high illuminance at the beginning of the test to stay stable in the afternoon zone. It is noteworthy that at the beginning of the experiment two illuminance spikes were introduced between the extreme values of very dark (night) and very high (midday) illuminance to simulate impossible situations in the form of outliers in the dataset, and in the middle of the experiment there was a spike train from high (afternoon) to normal (morning) illuminance at node *X1* that simulated the appearance of clouds in the sky during the $T_{ac}$.

**Table 2**
Datasets from Experiments 1 and 2, equivalent to a typical acquisition period of 60 min and 24 h per day for 15 days.

| Test scenario | Test profile description | Sampling period (s) | Total measurement time (s) | Sensor measurement dataset ($x^{(i)}$, $y^{(i)}$) |
|---|---|---|---|---|
| Experiment 1 | Combination of 3 temperature profiles: <br> - Indoor sudden cold-hot spikes from a normal baseline. <br> - Indoor cold drops from a normal temperature baseline. <br> - Stable normal outside temperature. | 10 | 3400 | 6460 |
| Experiment 2 | Combination of 3 temperature profiles: <br> - Indoor heat and cold-hot sawtooth. <br> - Indoor hot square waveforms from a cold baseline. <br> - Stable and cold-hot sawtooth outdoor. | 10 | 3400 | 6460 |

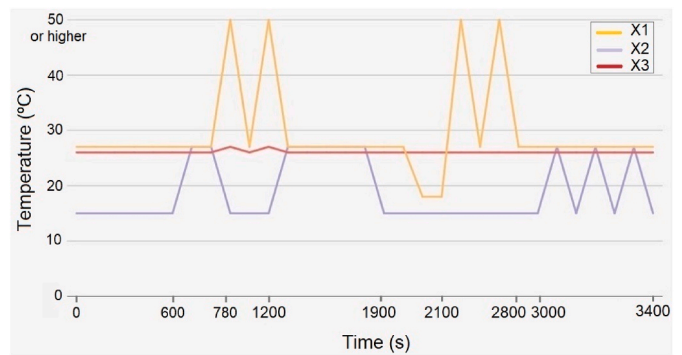

**Fig. 11.** Theoretical test profile of Experiment 1 for both IoT architectures.

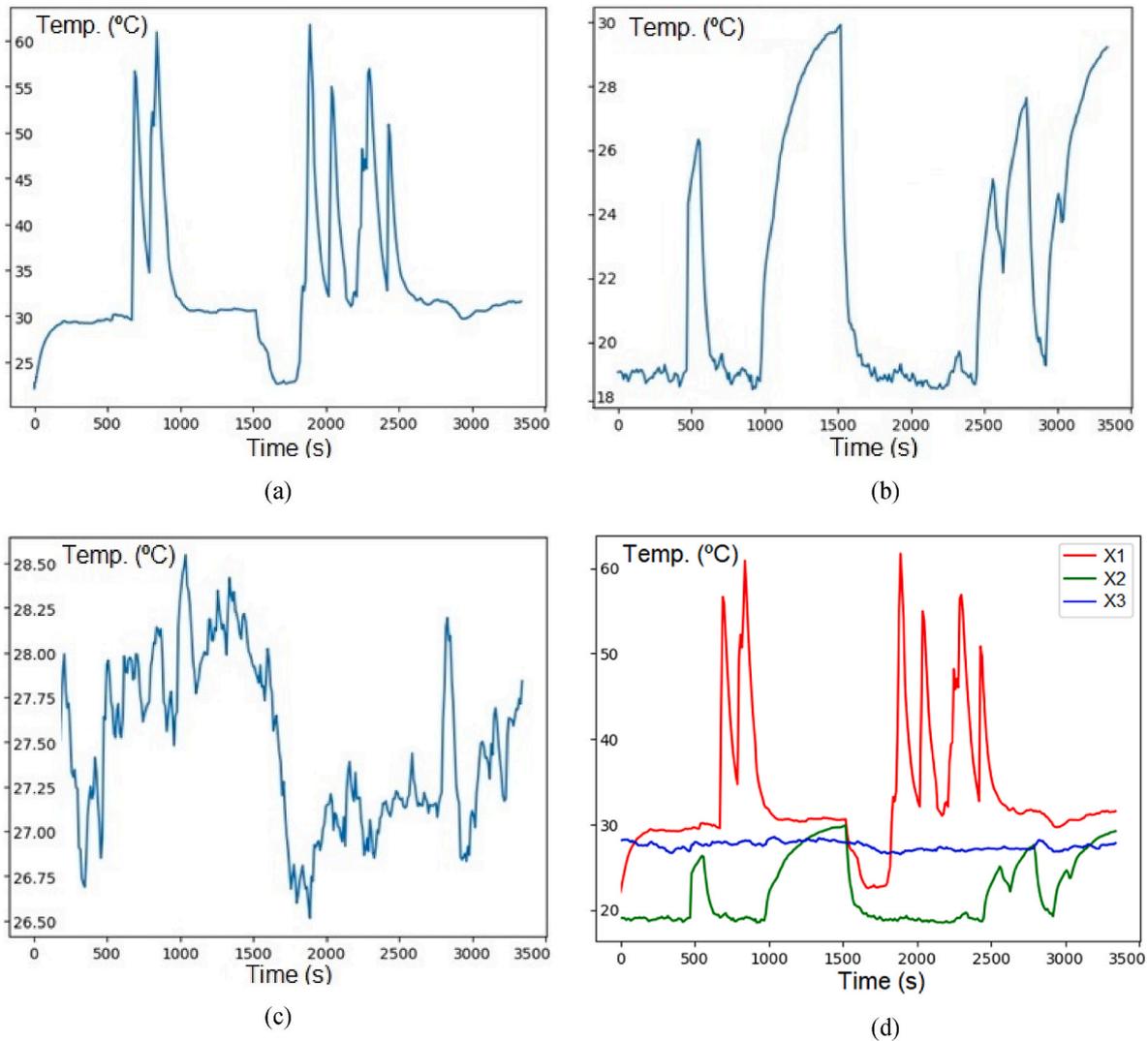

**Fig. 12.** Temperature data acquired for the Experiment 1 by each edge node: (a) *X1* node with indoor sudden spikes between extreme heat and cold from a normal temperature baseline, (b) *X2* node with indoor cold drops from a normal temperature baseline, (c) *X3* node with a stable outdoor temperature typical of summer in Spain, and (d) actual temperatures of all edge nodes.

**Table 3**
Digital and raw illuminance ranges.

| Scale | Illuminance digital range [a] | Illuminance raw value | Time slot of the day |
|---|---|---|---|
| 1 | 0–203 | Very Dark (VD) | Night |
| 2 | 204–408 | Dark (D) | Sunrise |
| 3 | 409–613 | Normal (N) | Morning |
| 4 | 614–818 | High (H) | Afternoon |
| 5 | 819–1023 | Very high (VH) | Midday |

[a] Digital steps of $(2^{10}/5)-1 \simeq 204$.

To evaluate the energy efficiency of the data acquisition in the Experiment 1, the instantaneous power consumption of each edge node based on an LPM and the comparison in consumption with respect to the edge node based on an HPM is shown in Fig. 14. In addition, Table 4 includes the average power consumption of these modules considering that each LPM had a RASPEX module attached whose consumption was low enough to hardly influence in the measurement of the LPM consumption.

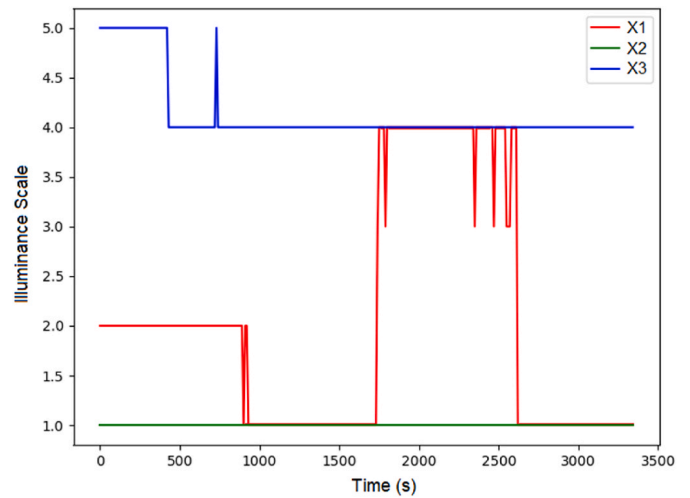

**Fig. 13.** Actual illuminance profile used in the experiment from the sensor fusion in the edge nodes.

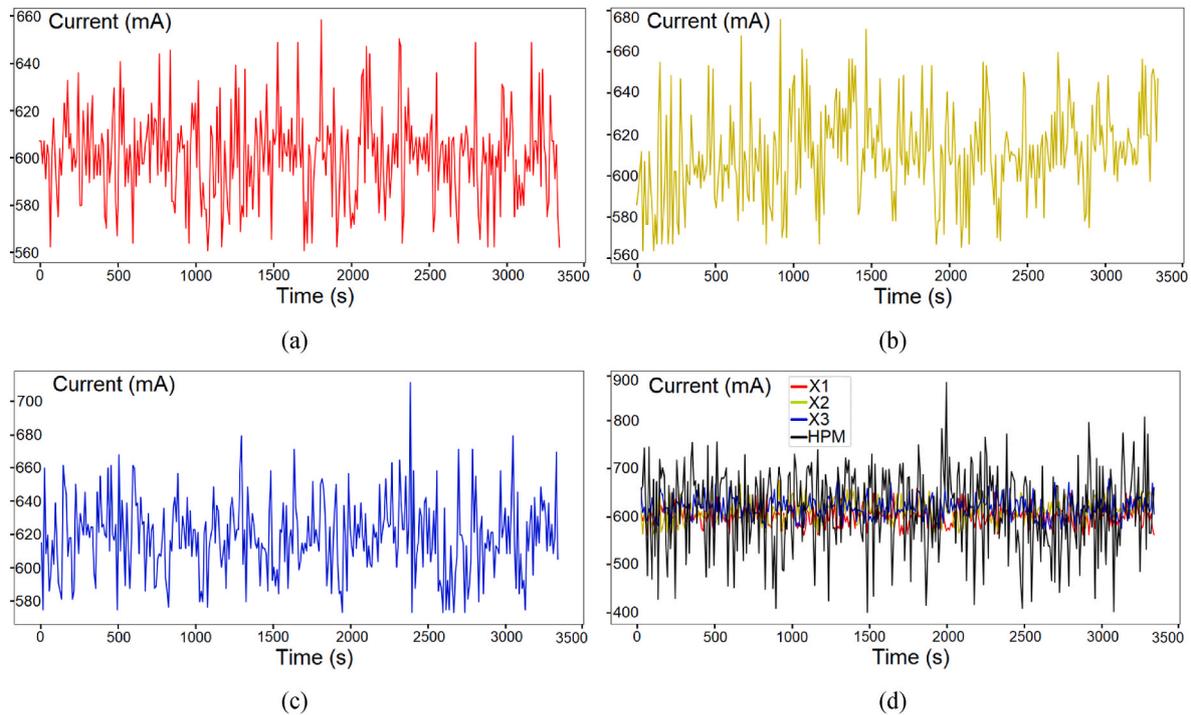

**Fig. 14.** Instantaneous direct current consumption of data acquisition in the Experiment 1 for LPM-based edge nodes: (a) *X*1 node, (b) *X*2 node, (c) *X*3 node, and (d) all nodes with respect to the HPM (GPU Nvidia Jetson Xavier NX).

**Table 4**
Average direct current and power consumption of architecture components during data acquisition.

| Architecture module | Power source voltage (V) | Average current consumption (mA) | Average power consumption (W) |
|---|---|---|---|
| *X*1 node (LPM-1 and RASPEX-1) | 5.1 | 600.14 | 3.06 |
| *X*2 node (LPM-2 and RASPEX-2) | 5.1 | 610.62 | 3.11 |
| *X*3 node (LPM-3 and RASPEX-3) | 5.1 | 618.32 | 3.15 |
| GPU (HPM) | 19 | 615.98 | 11.71 |

The Experiment 2 aimed to validate the robustness of parallel processing and try to increase the power consumption of processing to the worst case by considering a more complex model than those considered in Experiment 1, such as a 3-10-10-10-1 ML P-ANN. In this experiment, all edges nodes were unstable to obtain cold and hot temperature spikes, steps, and sawtooth bursts as shown in Fig. 15. The *X*1 node remained at a stable temperature for the first 840 s, and from this time until 1800 s, the manual air heater was used to heat the node to the maximum with a ripple of approximately 5 °C, then returned to its normal temperature and became excited by extreme spikes of heat and cold for 1000 s. As the *X*2 node was hanging from the ceiling of the room, it was cooled with the wall-mounted HVAC to obtain three temperature steps of about 500 s duration between the room temperature of 42 °C and the cooling down to 18 °C, which are oscillations typical in summer between night and midday. The *X*3 node was forced to acquire two sawtooth bursts between 18 °C and 42 °C again, these changes being more moderate than the *X*1 node. Both nodes were excited to achieve changes of more than 20 °C in 100 s using a manual air heater for rapid heating and ices cubes with a manual fan for rapid cooling, which was very useful to validate the robustness of the ML models against periodic and extreme

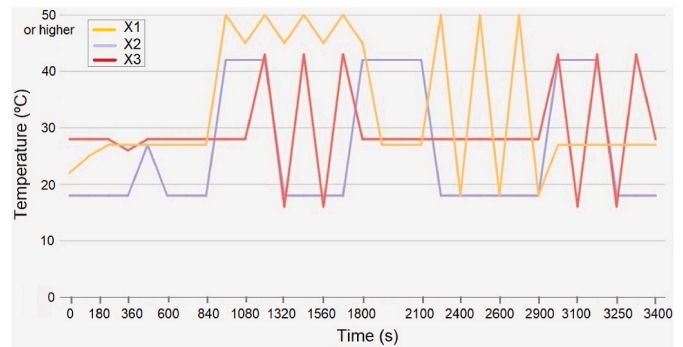

**Fig. 15.** Theoretical test profile of Experiment 2 only for the distributed parallel architecture for worst-case validation (stress testing).

variations.

Based on the theoretical test profile, the actual temperature data acquired by each edge node for the Experiment 2 is shown in Fig. 16. One important feature of the test profile in this experiment was the sharp and sudden spikes from cold to heat for checking the dynamic performance of the ML model. Another notable feature of the test was the steps with a heat-relevant time of about 200 s from a cold baseline of about 500 s duration to achieve at least two edge nodes in extreme cold or heat for a considerable time.

### 4.2. Training and testing of models

The procedures of ANN training, validation, and testing were performed on measurements obtained from the two experiments described above, essentially to estimate the temperature by time slot of the day. The dataset measurements were collected at an acquisition rate of 0.1 Hz and rearranged randomly prior to allocation to a specific subset. It provided symmetry breaking and generalization error reduction that substantially avoided the influence of overfitting (high variance) and underfitting (high bias) problems. As mentioned, in this work, the total number of ground truth measured data for training, validation, and testing was equal to 6460 for each experiment, which may contain outliers and missing values than may skew the results (avoided with the pre-processing stage shown in the general estimation methodology). Because proper data division is relevant as the results may be biased depending on how the data is split, the measured values were divided into fractions. As already indicated, the data division was adopted as a 70% fraction of the ground truth data for training and a 30% fraction for testing to assess the generalizability property and provide an independent measure of performance after training. It should be noted that the training dataset was selected small enough to reduce the processing load of the training process with an adequate learning rate, allowing the possibility of training offline or online in LPMs and HPM to have a proper balance between energy efficiency and estimation performance.

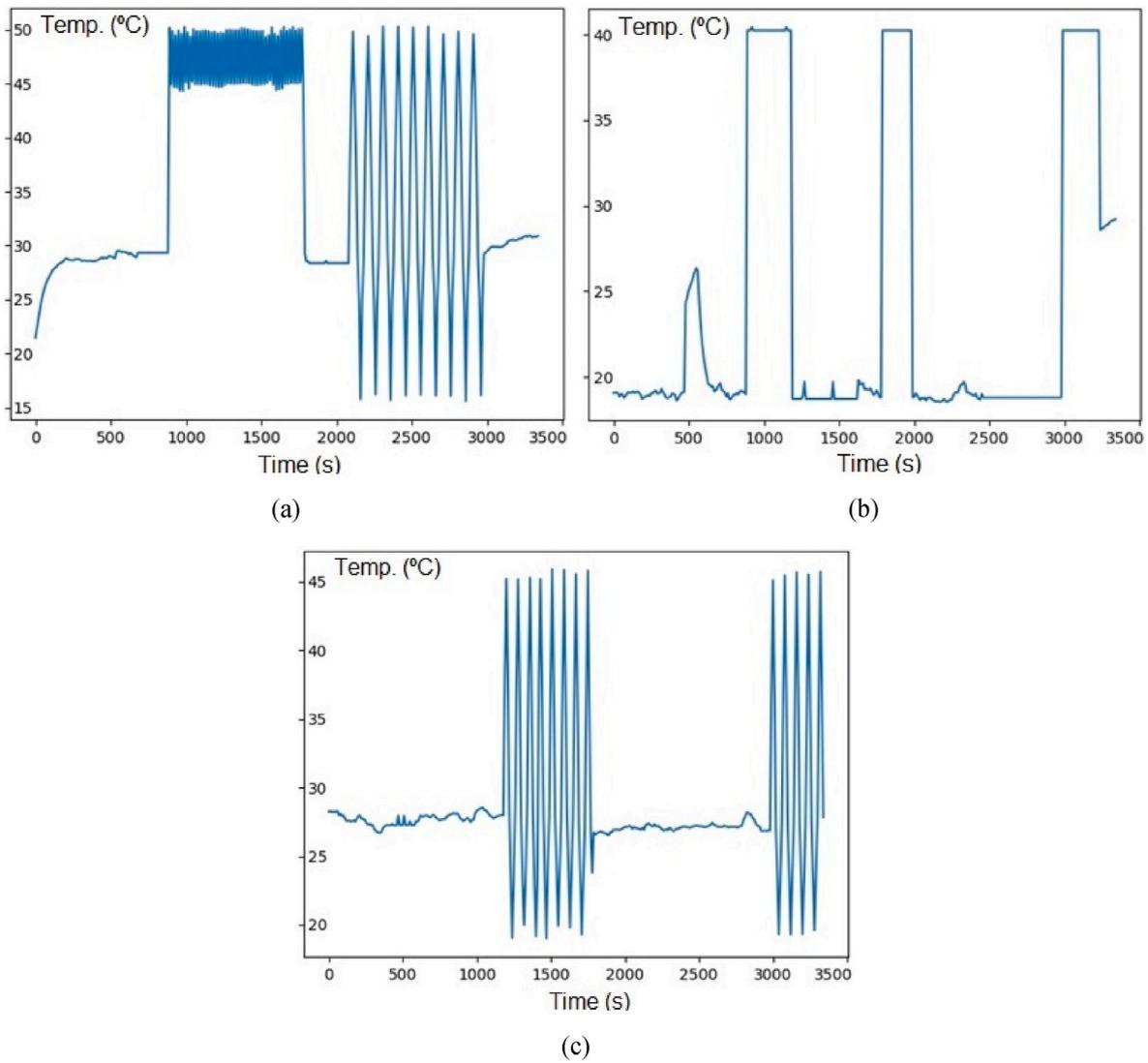

**Fig. 16.** Temperature data acquired for the Experiment 2 by each edge node: (a) *X1* node with indoor extreme heat and a sawtooth waveform of extreme heat and cold, (b) *X2* node with indoor hot square waveforms from a cold baseline, and (c) *X3* node with a stable outdoor temperature (summer in Spain) and sawtooth waveforms of extreme heat and cold.

In the first phase of the estimation methodology proposed in Section 3, the sensor data was acquired and recorded in databases to obtain the model inputs and then the criteria to label expected and estimated outputs were stablished. The experiments were carried out in an indoor environment with adequate dimensions and free air circulation, which allowed the gradients of the air quality parameters to be evenly distributed throughout the environment. For this reason, in the experiments based on the temperature estimation, the arithmetic mean of the instantaneous sensor measurements and the existence condition of this value within the range [26, 31] °C were used as the expected output as the suitable temperature in an indoor environment (Yan et al., 2019). These criteria are summarised as follows for each training example $i$:

$$y^{(i)} = \begin{cases} \frac{x_1^{(i)} + x_2^{(i)} + x_3^{(i)}}{3} & \text{if } y^{(i)} \in [26, 31]\,°C \rightarrow y_{label}^{(i)} = 1\,(Valid, True) \\ 26\,°C & \text{if } y^{(i)} < 26\,°C \rightarrow y_{label}^{(i)} = 0\,(Invalid, False) \\ 31\,°C & \text{if } y^{(i)} > 31\,°C \rightarrow y_{label}^{(i)} = 0\,(Invalid, False) \end{cases}$$
(19)

where $y^i$ is the expected output for the training example $i$, $y_{label}^{(i)}$ is the label assigned to the training example $i$, and $x_1^i$, $x_2^i$, and $x_3^i$ are the instantaneous temperature values from edge nodes $X1$, $X2$, and $X3$, respectively.

The application of this labelling criteria to obtain the training examples $(x^{(i)}, y^{(i)})$ in the Experiment 1 is shown in Fig. 17.

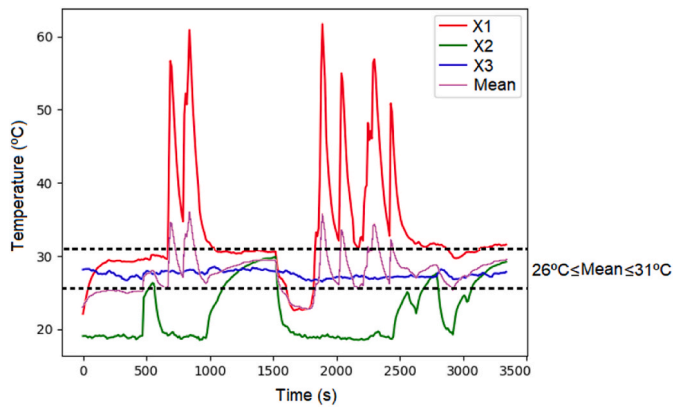

**Fig. 17.** Actual temperatures of all edge nodes with labelling criteria for data training in the Experiment 1.

*4.3. Discussion of estimation performance and relationship to energy efficiency*

The estimation performance against the energy efficiency was the ratio that was attempted to be increased to select an estimation method based on a specific edge IoT architecture. As starting point, the learning curves for all topologies of MLP-ANN were checked to assess the learning rate, verify that the training loss decreased to a point of stability, and identify underfit problems, as small datasets were used in the training to reduce data processing and therefore also power consumption. Some examples of learning curves for 1000 training epochs are shown in Fig. B.3 of Appendix B for ANN topologies with different TA deployed in the centralised architecture, such as the 3-2-2-1 topology with a low TA of 0.57 (Fig. B.3-a). It illustrates how the loss factor takes a bit longer to decrease as the number of epochs increases, and can be compared to the 3-6-6-1 topology with a higher TA of 0.99 (Fig. B.3-b) and a more drastic loss reduction. Fig. B.4 of Appendix B shows the learning curve for 4000 training epochs of the five-layer MLP-ANN with a 3–10-10-10-1 topology deployed in the distributed parallel architecture. In this curve, the loss factor is not close to zero until epoch 100 is reached, mainly due to the slightly higher training complexity of the ANN implemented in this architecture. This is because the trained model weights are distributed among the nodes and aggregated in the master node until the TA is low enough on this node.

As commented in Section 3, metrics based on the confusion matrix was used to evaluate the estimation performance. In this matrix, $y_{label}^{(i)}$ was used in one dimension with the labelling criteria of Equation (16) and the estimated output $\hat{y}^{(i)}$ was used in the other dimension using a tolerance for considering an estimation as valid. In the presented experiments, the temperature tolerance was 0.01 °C, which is at least equal to the sensor resolution according to Table A.2 of Appendix A. The labelling criteria of the estimated output for each testing example $i$ was:

$$\left|y^{(i)} - \hat{y}^{(i)}\right| = \begin{cases} Value \leq 0.01\ °C & \rightarrow \hat{y}_{label}^{(i)} = 1\ (Valid, True) \\ Value > 0.01\ °C & \rightarrow \hat{y}_{label}^{(i)} = 0\ (Invalid, False) \end{cases} \quad (20)$$

where $\hat{y}_{label}^{(i)}$ is the label assigned to the estimated output of the testing example $i$.

The performance results of temperature tests for the Experiment 1 in the centralised architecture are summarised in Table 5. In this experiment, ANN models were trained and tested centrally in the HPM, such as four-layer MLP-ANNs with topologies of different complexity in the hidden layers and trained with 500 and 1000 training epochs, obtaining a TA of at least 0.5. The estimation performance was slightly better for 1000 epochs, but the power consumption related to processing load was also higher with respect to 500 epochs. The most adequate balance between estimation performance and power consumption was obtained in the case of ANN topology for 500 training epochs and medium complexity (3-6-6-1) with a F-score of 0.96 after testing and an average consumption of 12.82 W. In this case, a RMSE of 2.82 and a MAE of 1.63 were obtained. For the more complex topology (3-50-50-1), a poor estimation performance is obtained possibly due to the need for a larger training set, which together with a higher consumption caused it to be discarded as a model. The simplest topology (3-2-2-1) is not selected owing to low performance estimation results, regardless of the training epochs and consumption. Meanwhile, there is no evidence of overfitting because the accuracy of the ML models is not significantly higher on the training dataset than on the test dataset in the selected ANN topology, but it is slightly higher in the simplest topology (3-2-2-1).

The performance results of temperature tests for the Experiment 1 in the distributed parallel architecture are summarised in Table 6. Based on the fact that in the centralised architecture the MLP-ANN 3-6-6-1 topology was selected for its performance and consumption, a somewhat more complex topology was implemented to try to worsen these two factors, nearly doubling the number of hidden neurons, adding a new hidden layer, and increasing the training time. That is why a five-layer MLP-ANN with a 3-10-10-10-1 topology was trained for 4000 epochs and tested sequentially and in parallel using in LPM devices with only one processing process per module. The results with sequential and parallel execution were similar for training with a TA of 0.95, but slightly better estimation performance with a F-score of 0.97 after testing and an average power consumption per node of 2.69 W were obtained. It provided a total consumption of 8.09 W, slightly lower than that obtained in sequential processing, but providing better scalability. In this case, an RMSE of 2.44 and a MAE of 1.45 were obtained. In addition, the parallel processing reached near super acceleration with a speedup of 3, close to the theoretical value since three distributed processes were used, and an efficiency of 0.98. In terms of parallel processing, an efficiency close to 1 means that the distribution of work was balanced among the three processors (three LPMs), so the system distributed the workload practically ideally. These results confirmed that barely better estimation performance was achieved in the distributed architecture compared to the centralised one with the same TA and a more complex model executed (MLP-ANN with one more hidden layer and twice more neurons in this layer), but needing four times more training epochs to achieve this performance. However, a reduction in power consumption close to 37% was obtained in the distributed architecture despite executing a more complex model and using more training time, due in part to the fact that when the slaves sent their weights to the master, they stopped CPU usage (goes into idle mode), so from this instant the power consumption of the slaves was minimal.

**Table 5**
Performance measurements of temperature tests for the Experiment 1 in the centralised architecture.

| Estimator model | Topology | TA | Precision | Recall | F-score | Accuracy | Average power consumption (W)[a] |
|---|---|---|---|---|---|---|---|
| Four-layer MLP-ANN (500 training epochs) | 3-2-2-1 | 0.50 | 0.69 | 0.10 | 0.17 | 0.12 | 12.71 |
|  | 3-6-6-1 | 0.93 | 0.93 | 0.99 | 0.96 | 0.92 | 12.82 |
|  | 3-50-50-1 | 0.73 | 0.93 | 0.97 | 0.95 | 0.90 | 13.49 |
| Four-layer MLP-ANN (1000 training epochs) | 3-2-2-1 | 0.57 | 0.84 | 0.28 | 0.42 | 0.28 | 12.04 |
|  | 3-6-6-1 | 0.99 | 0.93 | 0.93 | 0.93 | 0.86 | 12.56 |
|  | 3-50-50-1 | 0.59 | 0.94 | 0.89 | 0.92 | 0.85 | 12.66 |

[a] Power consumption of the processing module HPM (GPU Nvidia Jetson Xavier NX).

**Table 6**
Performance measurements of temperature tests for the Experiment 1 in the distributed parallel architecture.

| Estimator model | ML model execution | Parallel metrics | TA | F-score | Accuracy | Average power consumption (W)[c] | Total power consumption (W)[d] |
|---|---|---|---|---|---|---|---|
| Five-layer MLP-ANN with a 3-10-10-10-1 topology (4000 training epochs) | Sequential[a] Parallel[b] (three processes) | $T_S$ = 736.46 s $T_p$ = 249.89 s Speedup = 2.95 Efficiency = 0.98 | 0.95 0.95 | 0.95 0.97 | 0.93 0.95 | 2.71 2.69 | 8.11 8.09 |

[a] Model executed only in the master node (X1) while the slave nodes (X2 and X3) acquired sensor data and sent it to the master.
[b] Model executed in 3 nodes simultaneously and finally only the master performed processing with model weights computed in slaves.
[c] Power consumption per each processing module LPM (Raspberry Pi 3 B) with a CPU usage peak of 34%.
[d] Aggregated power consumption of three processing modules LPM.

Once the Experiment 1 was finished and evidencing that the distributed parallel architecture presented a higher performance-consumption ratio, a more demanding validation in real-time was carried out with the Experiment 2 using the same MLP-ANN topology. Unlike Experiment 1, in which the data was first acquired and stored and then the training and testing of the model was carried out, in the Experiment 2 the training and testing were accomplished in real time to verify the influence on the performance of the architecture components. Some of the factors that could affect performance were latencies from communication protocols such as MQTT and IEEE 802.11 b/g/n, overhead of the proposed data frame format and the encapsulation and decapsulation processes, delays due to software layers for data acquisition and processing, delays in database reads and writes and memory consistency issues, etc. The results obtained are shown in Table 7 for each iteration in real-time. The TA was 0.99 and the F-score was close to 0.95 throughout the experiment, which was a similar estimation performance to the Experiment 1 with an analogous power consumption of around 8 W. It demonstrated that the implemented architecture with its hardware components and software layers were correctly integrated and could be used to implement edge processing nodes in IoT applications that require real-time forecasts.

As auxiliary results, the estimation of the illuminance was also implemented to forecast the part of the day in which the temperature was being analysed. This estimation was also implemented with some of the MLP-ANN models used for temperature forecasting in the centralised architecture. Only the simplest MLP-ANN topologies (3-6-6-1 and 3-2-2-1) were trained for 1000 epochs and tested because illuminance data was simple (only structured in five scales) and it was important to have a moderate consumption for this auxiliary forecast, as shown in Table 8. The results revealed that the estimation performance with a F-score of 0.95 and the power consumption close to 12 W were very similar to those obtained in the temperature forecast, and it was sufficient to use the MLP-ANN with 3-2-2-1 topology instead of the 3-6-6-1 used for temperature.

Fig. 18 shows the test results of temperature estimation for the first 1000 s (100 samples) of the Experiment 1 using the selected ML model, such as the MLP-ANN with 3-6-6-1 topology. Considering the estimation methodology designed, the test dataset was obtained after randomly reordering the examples $(x^{(i)}, y^{(i)})$, so the test profile has an unknown shape that does not match the input dataset. As expected, the estimation curve has roughly centred on the stable temperature data from node X3, and spikes are obtained as a function of the temperature values towards hot or cold from the unstable data sources (X1 and X2 nodes). It is observed that there were no outliers in the estimation curve and the forecasts represented the mean value among all the data form the sensor nodes, which was appropriate in terms of uniform temperature distribution based on the temperature gradients of the measured indoor areas. As additional validation of the fit of the estimation with respect to the expected values, Fig. 19 shows the comparison of both values for the first 1000 s of the Experiment 1, verifying that the expected values are concentrated around the straight line of estimation with low data dispersion and negligible errors, which agrees with the optimal results obtained in the test metrics based on the confusion matrix, such as F-score and Accuracy. The F-score was preferentially considered as the combined evaluation metric, since it symmetrically represents both Precision and Recall in one metric. The results were 0.96 for the F-score and, for reference only, 0.92 for the Accuracy, supporting that the classifier was extremely precise and accurate, and omitted a relatively small number of examples.

Although in the present work the power consumption of the network is considered, it could be improved since it has some influence in the energy efficiency of the proposed ML estimation methods. This aspect is considered with the use of data frames with low overhead and easy encapsulation and decapsulation processes, some sensor nodes are put in sleep or idle modes for some time while others are active (for example, master and slave nodes in parallel processing), etc. However, the power consumption could be improved by considering another type of network, such as Zigbee, and by designing an energy saving WSN network with low latency and high throughput where key parameters of the Zigbee MAC layer may be determined by using a multi-criteria decision making, such as VIKOR and TOPSIS (Tsang et al., 2016). Another improvement could be to develop a health monitoring service to assess the energy efficiency of multi-hop sensors, as they are highly efficient because they refrain from long-distance data transmission, reducing energy by setting the default transmission power with the help of the channel status and routing tree (Alam et al., 2017).

### 4.4. Comparison to related research

The comparison of the proposed method with related research works is discussed in detail below. Similar solutions based on IoT architectures mainly, and other simpler approaches such as simulations or commercial data loggers, are considered to compare their performance estimation with respect to the methods proposed in the present work. In the comparison, similar estimated parameters and test conditions, including both indoor and outdoor environments, are considered as far as possible to gain a broader perspective of the architectural solutions offered. Table 9 includes related research works for indoor environments using edge-IoT and cloud-IoT architectures, or commercial data loggers. In addition, Table 10 lists similar IoT architectures for outdoor environments, including fog-IoT, and, for reference only, other studies for indoor environments based on simulations or surveys. In the performance comparison, the metrics used in the developed methods are preferably considered, such as F-score, Accuracy, RMSE and MAE, in addition to others used in related works such as the coefficient of determination ($R^2$) to evaluate the association between actual and predicted values (best prediction closest to 1) (Moursi et al., 2021) and the mean absolute percentage error (MAPE), which is suitable for comparing the prediction accuracy between different pollutants (less than 20% represents a decent prediction) (Liu et al., 2023).

**Table 7**
Performance measurements of real-time temperature tests for the Experiment 2 in the distributed parallel architecture.

| Estimator model | Model execution | Iteration number | TA | F-score | Accuracy | Average power consumption (W)[a] | Total power consumption (W)[b] |
|---|---|---|---|---|---|---|---|
| Five-layer MLP-ANN with a 3-10-10-10-1 topology | Parallel | 1st 2nd and following | 0.99 0.99 | 0.96 0.94 | 0.94 0.92 | 2.70 (529.41 mA) 2.68 (525.49 mA) | 8.10 8.04 |

[a] Power consumption per each processing module LPM (Raspberry Pi 3 B) with a CPU usage peak of 36%.
[b] Aggregated power consumption of three processing modules LPM.

**Table 8**
Performance measurements of experimental illuminance tests in the centralised IoT architecture.

| Estimator model | Topology | TA | Precision | Recall | F-score | Accuracy | Average power consumption (W)[a] |
|---|---|---|---|---|---|---|---|
| Four-layer MLP-ANN (1000 epochs training) | 3-2-2-1 | 0.96 | 0.94 | 0.97 | 0.95 | 0.92 | 12.01 |
|  | 3-6-6-1 | 0.98 | 0.95 | 0.99 | 0.97 | 0.94 | 12.12 |

[a] Power consumption of the processing module HPM (GPU Nvidia Jetson Xavier NX).

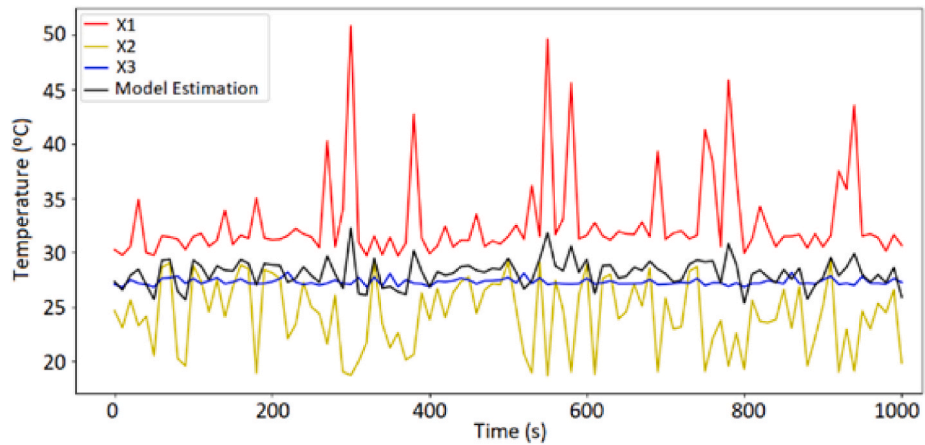

**Fig. 18.** Test profile of temperature estimation versus measurements of edge nodes ($X1$, $X2$ and $X3$) for the Experiment 1 using the MLP-ANN 3-6-6-1 topology.

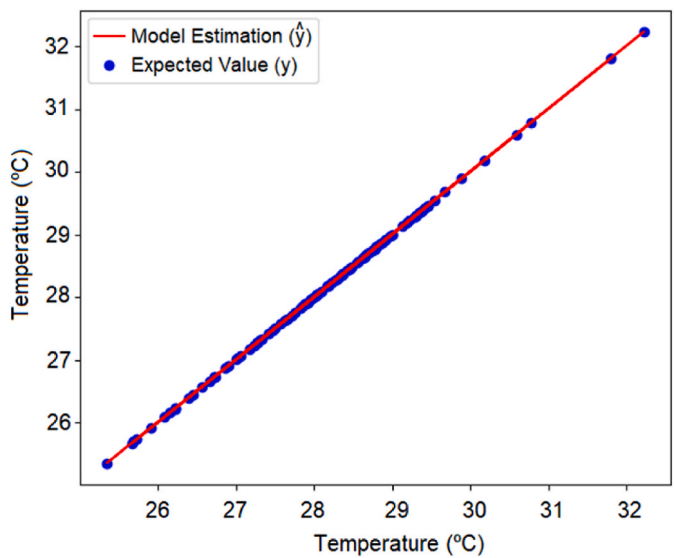

**Fig. 19.** Test results of temperature estimation versus expected outputs using the MLP-ANN 3-6-6-1 topology for the first 1000 s of the Experiment 1.

In addition, not only the MLP-ANN estimation models are included in the comparison of performance, but also other ML models are considered for a wider perspective, such as SVM, RF, Recurrent ANN (R-ANN) Long Short-term Memory (LTSM), Linear Regression (LR), Decision Tree (DT), Gradient Boost Decision Tree (GBDT), Extreme Gradient Boost (XGB) DT, Naive Bayes (NB), k-Nearest Neighbours (kNN), Least-squares Support Vector Machine (LSSVM) combined with Kernel Density Estimation (KDE), the hybrid model of Extreme Learning Machine (ELM) with Gray Wolf Optimizer (GWO), and Non-linear AutoRegression with eXogenous input (NARX). Among the most widely used in the estimation of IEQ parameters in indoor and outdoor environments, not only MLP-ANN but also SVM, RF, kNN, and DT stand out.

Firstly, the comparison of the estimation performance results of the selected method, based on distributed parallel processing, with respect to similar edge-IoT architectures for air quality parameters in indoor environments, such as temperature, RH and $CO_2$ is analysed. It is found that although the present work obtains a F-score/Accuracy higher to 0.95, a RMSE of 2.44 and a MAE of 1.45, some of the models implemented in edge-IoT and cloud-IoT architectures obtain an RMSE of 5–20 for MLP-ANN/SVM/RF models (Troncoso-Pastoriza et al., 2022), RMSE of 24 and F-score/Accuracy of 0.99 for the R-ANN LTSM model (Moursi et al., 2021; Mumtaz et al., 2021), RMSE of 24 for RF/GBDT models (Moursi et al., 2021), RMSE of 9.45 and MAE of 11.97 for the SVM model (Almalawi et al., 2022), and Accuracy of 0.76–0.95 for SVM/RF/DT models (Shah et al., 2020). In addition, with simulations on a single computer, RMSE results of 0.46–0.88 for a four-layer MLP-ANN model (Cho and Moon, 2022), RMSE of 15 for hybrid ELM with GWO model (Hou et al., 2022), and RMSE 0.85–0.87 for RF/XGB models (Tang et al., 2022). Also, an Accuracy of 0.82 for models MLP-ANN/SVM/RF (MLP with various topologies of hidden layers studied) is obtained using IAQ surveys to collect environmental data (Wong et al., 2022). These data show that the proposed method with a distributed parallel architecture

**Table 9**

Comparison performance of ML models and architectures regarding related research in air quality estimation (part 1/2).

| Reference | ML estimation model | Performance results | System architecture, sensors and deployed software | Estimated parameters and test conditions |
|---|---|---|---|---|
| **Presented work** | MLP-ANN *Distributed parallel: ANN 3-10-10-10-1 *Centralised: ANN 3-6-6-1 | F-score: 0.97 Accuracy: 0.95 RMSE: 2.44 MAE: 1.45 Energy: 8.09 W | *Architecture (selected)*: **edge-IoT distributed parallel computing** *Sensors*: SVM30 (Temp., RH, gases), Mikroe-1396 (current), and NSL-19M51(illuminance). *Software*: Python and C libraries. MariaDB database. | Temp., RH, $CO_2$, VOC, illuminance, and power consumption of nodes. *Conditions (indoor)*: collect data in 10s. Dataset of 6460 samples (42 $m^2$). |
| Troncoso-Pastoriza et al. (2022) | MLP-ANN SVM RF LR | RMSE: 6–16 RMSE: 6–20 RMSE: 5–10 RMSE: 6–21 *Training times lower than 152s. | *Architecture*: **edge-IoT computing** Raspberry Pi Zero as core of Wi-Fi sensor network via AMQP protocol with RabbitMQ client/server message brokers. *Sensors*: SHT31-D (Temp./RH), MHZ-14 ($CO_2$), TSL2561 (illuminance), among others. *Software*: Python, influxDB, and Grafana. | Temp., RH, and $CO_2$ concentration per node. *Conditions (indoor)*: collect data in 10 min intervals. Dataset of 135000 samples (area of 825 $m^2$ and 6 m ceil height) |
| Almalawi et al. (2022) | SVM GBDT LR | RMSE: 9.45 MAE: 11.97 RMSE: 4.15 MAE: 5.21 RMSE: 18.78 MAE: 17.98 | *Architecture*: **cloud-IoT computing** Sensor data collected by a microcontroller and sends it to Raspberry Pi, linked to cloud server for data storing and processing. *Sensors*: Gas sensor and laser dust sensor. *Software*: N/A. Virtual machines and database-as-a-service. | $CO_2$, $SO_2$, $NO_2$, and particulate matters ($PM_{2.5}$, $PM_{10}$). *Conditions (indoor)*: forecast the Air Quality Index for the next 5 h in the area. |
| Moursi et al. (2021) | R-ANN LTSM RF GBDT NARX XGB DT | RMSE: ~24 $R^2$: ~0.9 (all ML models) | *Architecture*: **edge-IoT and cloud-IoT** Cloud server connected by LoRa to Raspberry Pi 4 that collects data from Arduino Uno/NodeMCU via Wi-Fi linked to sensors by MQTT (ZigBee) or $I^2C$/SPI. *Sensors*: MQ-7/MQ-135 gas, DHT-11/DHT-22 (Temp./RH) and PM sensors. *Software*: Python (sckit-learn), cloud Amazon EC2 (IaaS, CaaS), noSQL DB. | Temp., RH, $CO_2$, $NH_4$, $CH_4$, and particulate matters ($PM_1$, $PM_{2.5}$, $PM_{10}$). *Conditions (indoor)*: data collected in a period of 24 h to forecast the next hour. |
| Mumtaz et al. (2021) | R-ANN LTSM SVM NB kNN | F-score and Accuracy: 0.99 F-score and Accuracy: 0.96 (rest of models) *ANN: three-layer, 10 hidden neurons | *Architecture*: **cloud-IoT computing** Local/cloud server connected to a NodeMCU by Wi-Fi/4G that collects data from ATmega328p MCU linked to sensors. *Sensors*: DHT-11 (Temp., RH), HM3301 laser ($PM_{2.5}$) and MH-Z19 ($CO_2$). *Software*: Python, Android (web site) | Temp., RH, $CO_2$, CO, $NH_3$, $CH_4$, and particulate $PM_{2.5}$. *Conditions (indoor)*: collect data in 5 min interval at rate of 0.0034 Hz. Dataset of 36388 samples. |
| Shah et al. (2020) | SVM RF DT | Accuracy: 0.95 Accuracy: 0.82 Accuracy: 0.76 | *Architecture*: **edge-IoT computing** Nvidia Jetson Nano for ML processing and Arduino Uno to collect sensor data. *Sensors*: DHT22 (temp., RH), MQ-5 (CO), MQ-131 (ozone), MQ-136 ($SO_2$), MQ-135 ($NO_2$), and PMS5003 (particulate matters). *Software*: Linux OS and MySQL DB (Nvidia Jetson), PHP/HTML5 (web). | Temp., RH, CO, $NO_2$, $SO_2$, and particulate matters ($PM_{2.5}$ and $PM_{10}$). *Conditions (indoor)*: Data of sensor measurements (dataset details N/A) and government databases. |
| Liu et al. (2023) | LS-SVM combined with KDE | $R^2$: 0.9 MAPE: < 9% | *Architecture*: **single data logger** *Sensors*: a proton-transfer-reaction time-of-flight mass spectrometer. *Software*: N/A. | Two VOC parameters (6-MHO, 4-OPA). *Conditions (indoor)*: five-day period (classroom). |

Note: N/A = Not available, N/U=Not used.

obtains the following:

- F-score and Accuracy similar or slightly better than equivalent ML models, such as MLP-ANN, SVM, and RF.
- An average reduction of the RMSE close to 76% and an average reduction of the MAE close to 83%.
- Considering the MLP-ANN as the best model in the related research, a reduction of the RMSE close to 60% is obtained with respect to the best performing IoT architecture, which is an edge-IoT deployment with a Raspberry Pi as a central node and Advanced Message Queuing Protocol (AMQP) message protocol (Troncoso-Pastoriza et al., 2022). However, an increase in RMSE close to 175% with respect to a single-computer simulation of a four-layer MLP-ANN (Cho and Moon, 2022) is obtained, but it cannot be considered a real IoT deployment like the proposed methods in this article.

Secondly, the comparison of the estimation performance results of the selected method (F-score/Accuracy of 0.95, RMSE of 2.44, and MAE of 1.45) with respect to similar IoT architectures for air quality parameters in outdoor environments, such as temperature, RH, $CO_2$, CO, and VOC is analysed. It is found that some of the models implemented in cloud-IoT and fog-IoT architectures obtain an F-score/Accuracy of 0.94–0.99 for a MLP-ANN model and 0.71–0.99 for SVM/kNN/NB models in a city with sensors placed in buses (Ahmed et al., 2020), RMSE of 6–25 for SVM/RF/GBDT models in a small city covered by static and mobile sensors (Zhang and Woo, 2020), and RMSE of 5.36 for the SVM model and close to 0.1 for RF/kNN models in industrial environments (Garcia-Sanchez et al., 2022). These data show that the proposed method with a distributed parallel architecture obtains the following:

- F-score and Accuracy similar or slightly better than similar ML models, such as MLP-ANN, SVM, and kNN.
- An average reduction of the RMSE error close to 35% in which an architecture based on cloud-IoT computing using LoRaWAN (for sending data to cloud) and MQTT (for sensors connection) obtains great performance for temperature estimation in industrial environments (Garcia-Sanchez et al., 2022).

**Table 10**

Comparison performance of ML models and architectures regarding related research in air quality estimation (part 2/2).

| Reference | ML estimation model | Estimation performance | System architecture, sensors and deployed software | Estimated parameters and test conditions |
|---|---|---|---|---|
| **Presented work** | MLP-ANN<br>*Distributed parallel: ANN 3-10-10-10-1<br>*Centralised: ANN 3-6-6-1 | F-score: 0.97<br>Accuracy: 0.95<br>RMSE: 2.44<br>MAE: 1.45<br>Energy: 8.09 W | *Architecture (selected)*: edge-IoT distributed parallel computing<br>*Sensors*: SVM30 (Temp., RH, gases), Mikroe-1396 (current), and NSL-19M51 (illuminance).<br>*Software*: Python and C libraries. MariaDB database. | Temp., RH, $CO_2$, VOC, illuminance, and power consumption of edge nodes.<br>*Conditions (indoor)*: collect data in 10s. Dataset of 6460 samples (42 $m^2$ area). |
| Zhang and Woo (2020) | SVM, RF, GBDT | RMSE: 6–25<br>(all ML models, only for $PM_{2.5}$ and $PM_{10}$) | *Architecture*: cloud-IoT computing<br>Cloud server to Raspberry Pi 3B+ by LTE that collects sensor data. Arduino Mega send GPS data over VoLTE network.<br>*Sensors*: six nodes (static and mobile).<br>*Software*: DataPicker (server), Google Maps. | Temp., RH, $CO_2$, and particulate matters ($PM_{2.5}$ and $PM_{10}$).<br>*Conditions (outdoor-small city covered by sensors)*: collect data in 1 min intervals (few hours/day). Dataset of 4 days. |
| Ahmed et al. (2020) | MLP-ANN<br>SVM, kNN, NB<br>DT | Accuracy: 0.97<br>F-score: 0.94–0.99<br>Accuracy: 0.9–0.99<br>F-score: 0.71–0.99<br>Accuracy: 0.69<br>F-score: 0.7–0.86 | *Architecture*: **distributed fog-IoT computing**<br>Microcontroller collects sensor data and send it to fog layer by GSM/Wi-Fi.<br>*Sensors*: N/A<br>*Software*: iFogSim toolkit. | CO, $NO_2$, $O_3$, and $SO_2$.<br>*Conditions (outdoor-city with sensors on buses)*: Simulation. Collect data per hour. Dataset of 1.7 million samples. |
| Garcia-Sanchez et al. (2022) | SVM<br>RF<br>kNN<br>LR | RMSE: 5.36<br>RMSE: 0.07<br>RMSE: 0.15<br>RMSE: 3.05<br>*Results for Temperature | *Architecture*: **cloud-IoT computing**<br>Sensors connected by MQTT. Data sent to Linux server by LoRaWAN.<br>*Sensors*: N/A<br>*Software*: Python, influxDB database | Temp., RH, $CO_2$, $NO_2$, $O_3$, $SO_2$, and particulate ($PM_1$, $PM_{2.5}$, $PM_{10}$).<br>*Conditions (outdoor-industrial)*: Dataset of 33,000–70,000. |
| Tang et al. (2022) | RF<br>XGB DT | RMSE: 0.87<br>RMSE: 0.85<br>*Training: 5min, 1000 trees | *Architecture*: **single-computer simulation**<br>Intel Core i5-8350U (16 GB RAM) computer.<br>*Sensors*: N/A.<br>*Software*: Python (shap) and R (stats). | Temp., $CO_2$, illuminance, sound, and ventilation rate to analyse comfort.<br>*Conditions (indoor)*: simulation |
| Hou et al. (2022) | Hybrid ELM with GWO<br>*Three-layer ANN, 30 hidden neurons | RMSE: ~15<br>$R^2$: ~0.75 | *Architecture*: **single-computer simulation**<br>12-core Intel Xeon 6126<br>*Sensors*: UT332+ (Temp., RH) and Testo 535 ($CO_2$) data loggers<br>*Software*: Autodesk CFD (IAQ simulation), eQUEST (electricity use simulation). | Temp. and $CO_2$ (measured 1.1 m above ground).<br>*Conditions (indoor)*: simulation 2 h for 1000 iterations (50 h total). |
| Cho and Moon (2022) | MLP-ANN<br>*Four-layer ANN, selected from 12 models | RMSE: 0.46–0.88<br>$R^2$: ~0.99 (depends on the air quality parameter) | *Architecture*: **single-computer simulation**<br>*Sensors*: N/A<br>*Software*: CONTAM and Energyplus. Matlab (Deep Learning Toolbox). | $CO_2$ and particulate ($PM_{2.5}$, $PM_{10}$)<br>*Conditions (indoor-school)*: simulation. Dataset of 200000 samples. |
| Wong et al. (2022) | MLP-ANN<br>RF<br>SVM<br>kNN | Accuracy: ~0.82<br>*ANN: 1, 3, 4, and 6 hidden layers (100, 200 neurons/layer). | *Architecture*: **data collected from IAQ surveys** (525 air-conditioned offices). | $CO_2$, VOC, $NO_2$, CO, and respirable particulates.<br>*Conditions (indoor)*: 8 h continuous sampling. Dataset of 32 samples. |

Note: N/A = Not available, N/U=Not used.

Thirdly, in none of the analysed research works included in Tables 9 and 10, among many others, have been considered the processing times for training and testing the models, nor has the power consumption of the hardware devices used for model learning from sensor data. Only the works (Troncoso-Pastoriza et al., 2022) for edge-IoT computing and (Tang et al., 2022) for single-computer simulation have considered the training times of ML models, and (Mumtaz et al., 2021; Zhang and Woo, 2020) of cloud-IoT computing included slight comments on the power consumption of the architecture nodes without precise electrical data and not related to ML model computing or sensor data acquisition. On the other hand, in the methods presented in the article, based on two IoT architectures, the ML processing times have been considered in relation to power consumption (measured with DC current sensors) to select the appropriate estimation method that obtains the best balance between estimation performance and energy efficiency for responsible sustainable environments. For example, the acquisition period of sensor data was set to 10 s in the presented methods after several analyses to use the minimum level allowed because reducing the frequency of the samples could optimise the system power consumption over a pre-existing Wi-Fi infrastructure, which could be especially useful in the case of using the proposed architectures as battery-powered systems. In addition, in none of the research works studied have been found the use of parallel processing to distribute computing capacity among the nodes, which provides a greater control over power consumption and scalability of the architecture depending on the processing needs, being able to have nodes in active or inactive modes depending on the model complexity.

In summary, the proposed method significantly improves the average performance of the estimation of environmental parameters over edge-IoT and cloud-IoT architectures, surpassing even some architectures implemented with similar COTS resources. Specifically, it is achieved a slightly better performance of classification metrics (F-score and Accuracy) than ML models implemented in edge-IoT and cloud-IoT for indoor and outdoor environments. Furthermore, an average reduction of RMSE close to 76% and 35% are obtained for ML models deployed on indoor and outdoor IoT architectures, respectively, but perform worse than single-computer simulations that are not implemented on IoT architectures with real hardware and software. These results show that a possible refinement of the methods can be considered, such as the implementation of additional ML models such as RF and XGB DT that obtain a significant reduction of the RMSE close to 0.85 with relatively low training times for a single-computer simulation (Tang et al., 2022), and R-ANN LTSM models achieving F-scores and Accuracy close to 0.99 in cloud-IoT architectures (Mumtaz et al., 2021) whose communication latencies and energy efficiency could be improved with edge-IoT topologies.

## 5. Conclusions

This article presents a comparison of edge computing methodologies based on two IoT architectures for the estimation of environmental parameters. Centralised and distributed parallel architectures are proposed to implement some ML models, such as various MLP-ANN topologies, which are composed of edge nodes with different processing capability such as three LPM nodes based on ARM processors and one HPM node based on a GPU. Edge nodes are connected using a wireless network and data is exchanged via MQTT in the centralised architecture and via MPI in the distributed parallel, both using purpose-built lightweight data frames.

Although multiple environmental parameters have been measured, the results have been obtained based on the temperature estimation and, additionally, the illuminance estimation, reaching the following conclusions:

- The distributed parallel architecture obtains an estimation performance similar to the centralised one for the same experiments using small datasets with a F-score and an Accuracy close to 0.95. However, a parallel performance of 0.98 efficiency is attained (Speedup of 2.95, close to the theoretical value for three processors) which means the workload distributed among the processors is balanced, but executing a more complex MLP-ANN model with more hidden layers and neurons in a distributed way.
- The distributed parallel architecture obtains a reduction in power consumption close to 37% regarding the centralised approach although the executed ML model was more complex, and required more training time to reach a Training Accuracy of at least 0.95.

Regarding related research works, the comparison with equivalent IoT architectures for estimating environmental parameters shows the following:

- Compared to estimations in indoor environments, an F-score and Accuracy similar or slightly higher are obtained for analogous ML models, such as MLP-ANN, SVM, and RF. In addition, an average reduction of the RMSE close to 76% and an average reduction of the MAE close to 83% are achieved. The implementation of MLP-ANNs with an ARM processor (Raspberry Pi) and AMQP protocol is the best performing edge-IoT architecture found, achieving a RMSE reduction of close to 60%.
- Compared to estimations in outdoor environments, an F-score and Accuracy similar or slightly higher are obtained for analogous ML models, such as MLP-ANN, SVM, and kNN. In addition, an average reduction of the RMSE close to 35% is achieved regarding the best performing cloud-IoT architecture based on LoRaWAN and MQTT to exchange data with the cloud and sensors, respectively.
- For reference only, compared to single-computer simulations of analogous MLP-ANN models for indoor environments, an increase in RMSE close to 175% is obtained, but it cannot be considered a real IoT deployment like the proposed methods.
- Unlike the article presented, none of the research papers consider the processing times to train and test the ML models, nor the power consumption of the hardware to learn the model from sensor data and select an ML model with a proper balance between performance and energy efficiency.

In conclusion, this work presents the comparison of two edge-computing methods to show the significant improvement that a distributed parallel approach supposes in estimating performance against energy efficiency despite the greater complexity of ML model implementation, but providing better scalability at lower cost. In future research, the study could also cover other ML models, such as RF and XGB DT, use lightweight networking standards and messaging protocols such as ZigBee and CoAP, respectively, deploying a conventional cloud-IoT variant of the centralised architecture, and additionally, cloud intelligence locally on edge-IoT devices with cloud-native workloads, such as Microsoft Azure IoT Edge, using low-power COTS devices in order to compare results of parameter estimation and power consumption.

## CRediT authorship contribution statement

**Jose-Carlos Gamazo-Real:** Project administration, Funding acquisition, Conceptualization, Supervision, Investigation, Formal analysis, Methodology, Resources, Software, Validation, Data curation, Visualization, Writing - original draft, Writing - review & editing. **Raúl Torres Fernández:** Software, Validation, Data curation, Visualization. **Adrián Murillo Armas:** Software, Validation, Data curation, Visualization.

## Declaration of competing interest

The authors declare the following financial interests/personal relationships which may be considered as potential competing interests: Jose-Carlos Gamazo-Real reports article publishing charges was mainly provided by the Spanish national project with reference PID2020-118969RB-I00 ("Sistemas IoT Conscientes de la Sostenibilidad y Dirigidos por Comunidades Sociales") of the SYST research group from the Universidad Politécnica de Madrid (Madrid, Spain), and also partially provided by the CRUE Spanish Universities - CSIC Alliance. Jose-Carlos Gamazo-Real reports a relationship with the Universidad Politécnica de Madrid that includes: employment.

## Data availability

Other

## Acknowledgements

This work was mainly supported by the SYST (Systems and Software Technology Group) research group of the Universidad Politécnica de Madrid (Madrid, Spain) under the project with reference PID2020-118969RB-I00 and title "Sistemas IoT Conscientes de la Sostenibilidad y Dirigidos por Comunidades Sociales", identified with the acronym as "SIoTCom", of the Spanish national call 2020 "Proyectos de I + D + i" (within the framework of the "Programas Estatales de Generacion de Conocimiento y Fortalecimiento Científico y Tecnológico del Sistema de I+D+i y de I+D+i Orientada a los Retos de la Sociedad") in the area of Information Technologies and Communications, and the subarea of Computing Science and Computer Technology. It was also partially supported by the ETSI Sistemas Informáticos (ETSISI) of the Universidad Politécnica de Madrid within the program "Programa Propio ETSISI 2022" of the call for support the development of projects that require the implementation of a hardware subsystem. This work also received partial funding for open access charge of the Universidad Politécnica de Madrid.

## Appendix A

Additional details on the design of the experimental setup are included in this appendix to complement the content of Section 2. The main characteristics of LPM and HPM processing devices and sensor modules used in the experimental prototype are listed in Table A.1 and Table A.2, respectively. A complete description of the sensor data frame fields used in the communication between processing modules is included in Table A.3. The implementation of the software components was performed with the C and Python programming languages, specifically using the external libraries listed in Table A.4. In addition, Fig. A.1-a shows the specific data frame format used in the presented paper with a size of 49 bytes and Fig. A.1-b shows an example of a temperature measurement encapsulated using the IEEE 754 format (four bytes) in the "sensor measure" field of the data frame. Finally, Fig. A.2 shows a MariaDB database query access to display the complete list of sensor measurements.

**Table A.1**
Main characteristics of the processing modules in the experimental prototype.

| Module | Processor | Processing cores | RAM memory | Cache memory | WLAN interface |
|---|---|---|---|---|---|
| LPM | Broadcom BCM2837 64-bit CPU (1.2 GHz) | 4 ARM Cortex-A53 cores | 1 GB LPDDR2 RAM (900 MHz) | 32 kB L1 512 kB L2 | 802.11 n 2.4 GHz |
| HPM | Nvidia Carmel ARM v8.2 64-bit CPU (1.1 GHz) and Nvidia Volta GPU | 6-core CPU and 384 CUDA cores with 48 Tensor cores GPU | 8 GB LPDDR4 RAM (1866 MHz) | 6 MB L2 4 MB L3 | 802.11 ac 2.4–5 GHz |

**Table A.2**
Main characteristics of the sensor modules in the experimental prototype.

| Sensor module | Parameter | Accuracy tolerance | Resolution | Operating range | Unit |
|---|---|---|---|---|---|
| SENSAIR | Temperature | ±1 | 0.01 | (-20, 85) | °C |
| | Relative Humidity | ±5 | 0.01 | (0, 100) | %RH |
| | $CO_2$ | – | 9 ppm (Typ.) | (400, 60,000) | ppm[a] |
| | VOC | – | 6 ppb (Typ.) | (0, 60,000) | ppb[a] |
| SENSLUX | Illuminance | – | 0.097 (Typ.) | (10, 100) for (5, 100) kΩ | lux |
| SENSCUR | DC Current | – | 1.9 (Typ.) | (100, 2048) | mA |

[a] ppm: parts per million, ppb: parts per billion.

**Table A.3**
Fields of the data frame to encapsulate the sensor data.

| Data frame section | Size (byte) | Field | Field index (byte)[a] | Content type |
|---|---|---|---|---|
| Header | 5 | Type of frame | 0 | Integer: 0 = Sensor data, 1 = Control data |
| | | Device/Node ID | 1–3 | Unsigned integer: $0 - (2^{24}-1)$ |
| | | Number of sensors | 4 | Unsigned integer: 1–255 |
| Timestamp (MySQL format) | 6 | YYYY (year) | 5-6 (12 bits) | Unsigned integer: 1970–2100 |
| | | MM (month) | 6 (4 bits) | Unsigned integer: 1–12 |
| | | DD (day) | 7 (5 bits) | Unsigned integer: 1–31 |
| | | HH (hour) | 7-8 (5 bits) | Unsigned integer: 0–23 |
| | | MM (minutes) | 8 (6 bits) | Unsigned integer: 0–59 |
| | | SS (seconds) | 9 (6 bits) | Unsigned integer: 0–59 |
| | | MS (milliseconds) | 9-10 (10 bits) | Unsigned integer: 0–999 |
| Data | 5 | Sensor ID | 11 | 0: Temperature, 1: RH, 2: Illuminance, 3: $CO_2$, 6: Current consumption, 10: VOC |
| | | Sensor measure | 12–15 | Floating point in IEEE 754 format |
| Tail | 2 | Newline | $N$-2 | '\n' |
| | | Carriage return | $N$-1 | '\r' |

[a] Absolute position in bytes in the frame (starting at 0).

**Table A.4**
External C and Python libraries used in software implementation.

| Programming language | Libraries | Description and use |
|---|---|---|
| C language (gcc compiler 8.30) | Stdio.h | Standard input/output (I/O) for monitoring tasks |
| | Time.h | Timestamp implementation and delays setup |
| | WiringPi.h | Access to Raspberry general purpose I/O pins |
| | WiringPiI2C.h | Control of devices connected to Raspberry I$^2$C bus |
| | WiringPiI2C.h | Control of devices connected to Raspberry SPI bus |

(*continued on next page*)

**Table A.4** (*continued*)

| Programming language | Libraries | Description and use |
| --- | --- | --- |
| Python language (version 3.6.9, pip 22.0.3) | Mosquitto.h | Configuration of MQTT topics, subscribers, publishers, etc. |
| | Mysql.h | Queries and connection to MySQL databases in LPMs |
| | Svm30.h | Read and configuration of Sensirion SVM30 sensor module |
| | Tensorflow + Keras | Configuration of neural networks for aarch64 architectures |
| | Pandas | Data import and management for algorithm development |
| | Numpy | General operations between data structures |
| | Matplotlib | Representation of data in graphs |
| | Statsmodels | Calculation of statistical metrics |
| | MysqlDB | Queries and connection to MySQL databases in HPM |
| | MPI4py | Message passing interface directives for LPM data exchange |

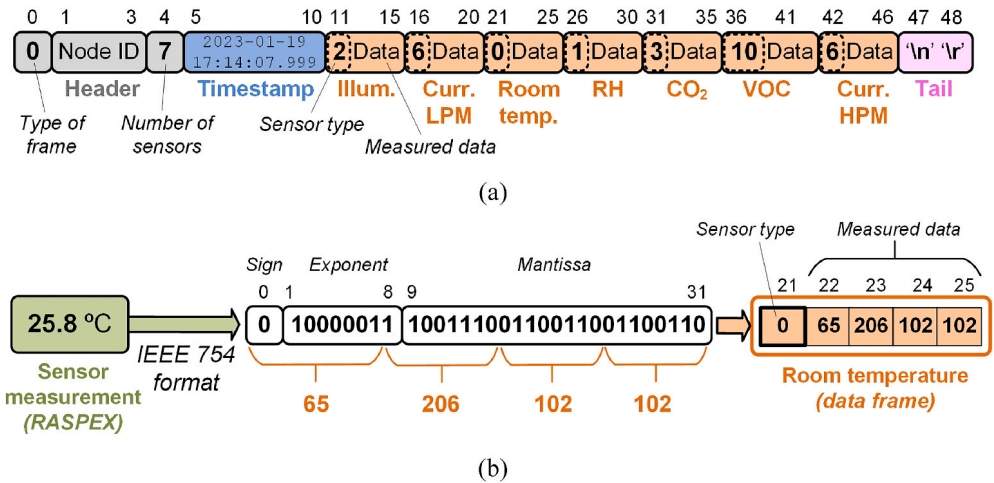

(a)

(b)

**Fig. A.1.** Fields of the data frame used in the presented paper: (a) complete data frame with all sensors and (b) formatting process in IEEE 754 of a temperature measurement until its encapsulation in the data frame.

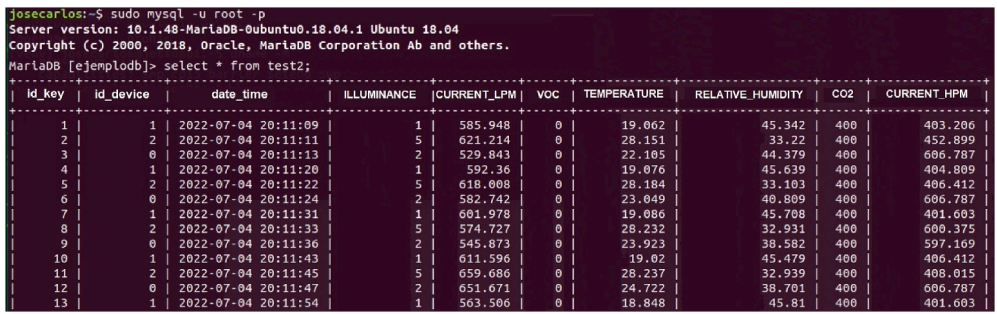

**Fig. A.2.** MariaDB database query access to display the complete list of sensor measurements.

**Appendix B**

Additional details on the experimental estimation methods and results are included in this appendix to complement the content of Sections 3 and 4. Fig. B.1 shows 3-6-6-2 and 3-50-50-1 ANN topologies used in the centralised architecture. The location photos of edge nodes $X2$ and $X3$ in the room for experimental tests are shown in Fig. B.2-a and Fig. B.2-b, respectively. Fig. B.3 shows some examples of learning curves for 1000 training epochs for ANN topologies with different TA deployed in the centralised architecture, such as a 3-2-2-1 topology (Fig. B.3-a) and a 3-6-6-1 topology (Fig. B.3-b). Fig. B.4 shows the learning curve for 4000 training epochs of the five-layer MLP-ANN with a 3–10-10-10-1 topology deployed in the distributed parallel architecture.

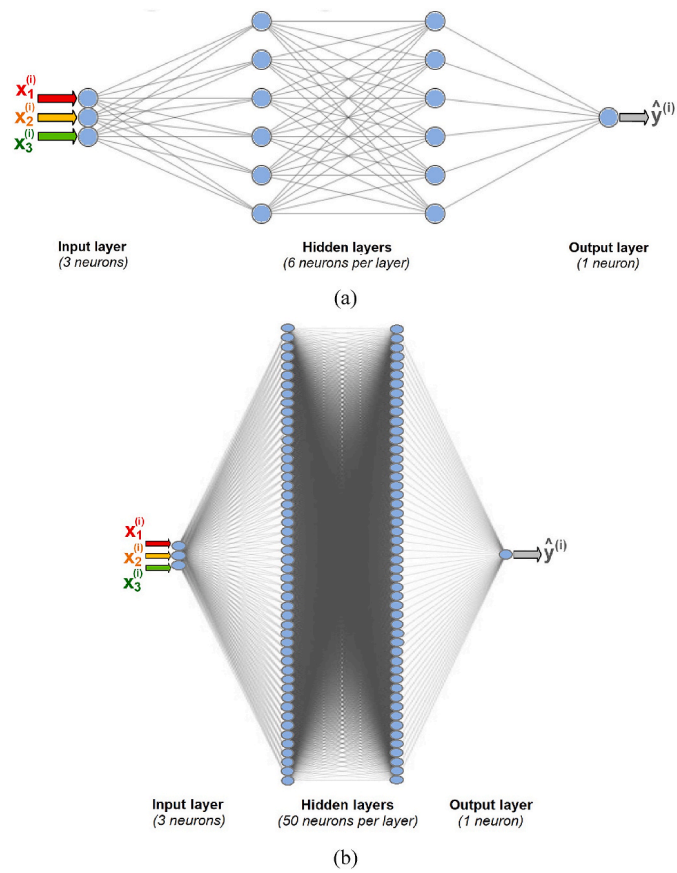

Fig. B.1. Implemented MLP-ANNs with different four-layer topologies: (a) 3-6-6-2 and (b) 3-50-50-1.

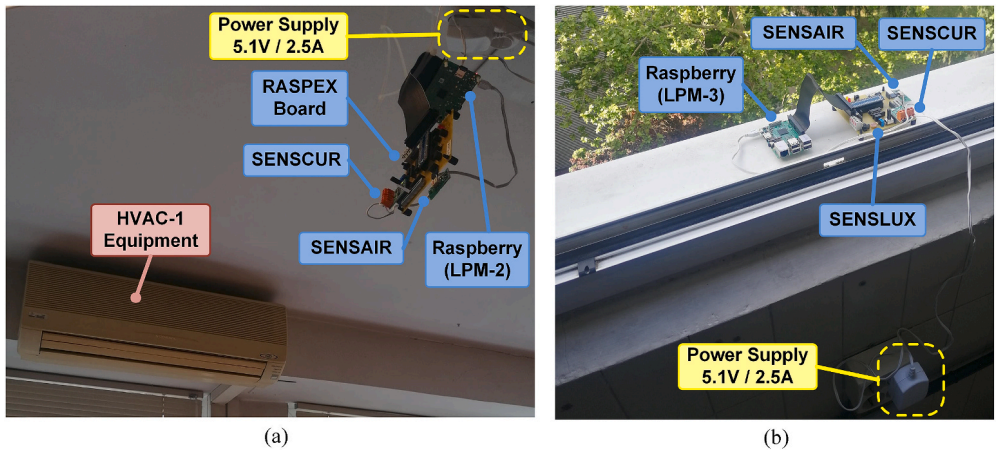

Fig. B.2. Indoor environment for experimental tests: (a) location photo of the *X*2 edge node and (b) location photo of the *X*3 edge node.

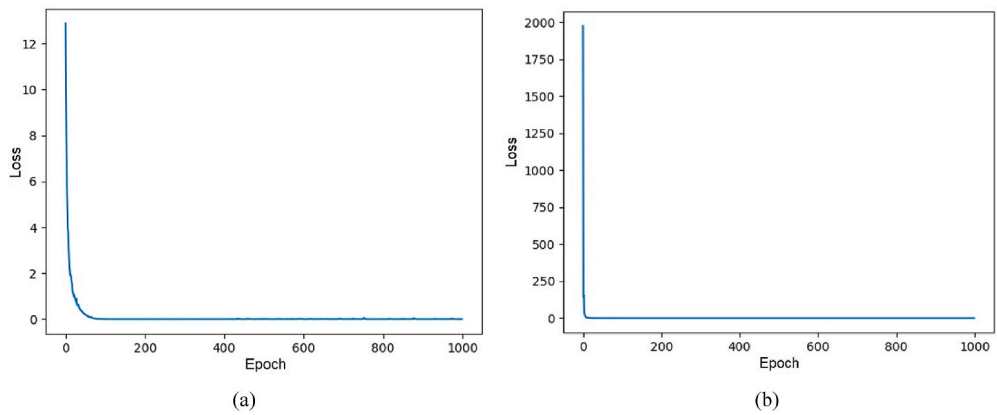

**Fig. B.3.** Learning curves in the centralised architecture for 1000 training epochs of four-layer MLP-ANNs with topologies: (a) 3-2-2-1 and (b) 3-6-6-1.

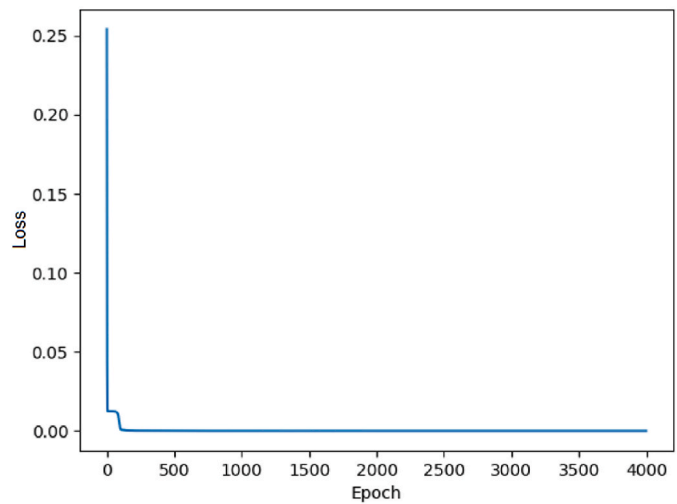

**Fig. B.4.** Learning curve in the distributed parallel architecture for 4000 training epochs of a five-layer MLP-ANN with a 3-10-10-10-1 topology.